\documentclass[acmsmall]{acmart}

\usepackage[utf8]{inputenc}
\usepackage{wrapfig}
\usepackage{graphicx}
\usepackage{subfig}
\usepackage{multicol}
\usepackage{multirow}

\usepackage{booktabs}
\usepackage{tikz}
\usepackage{pgfplots}

\usepackage{tcolorbox,fancyvrb,xcolor,tikz}

\usepackage{listings}

\usepackage{algorithm}
\usepackage{algorithmic}

\usepackage{xspace}

\usepackage{url}

\AtBeginDocument{%
  \providecommand\BibTeX{{%
    \normalfont B\kern-0.5em{\scshape i\kern-0.25em b}\kern-0.8em\TeX}}}

\colorlet{punct}{red!60!black}
\definecolor{background}{HTML}{EEEEEE}
\definecolor{delim}{RGB}{20,105,176}
\colorlet{numb}{magenta!60!black}

\newcommand{\ubf}[1]{\textbf{\mathversion{bold}#1}}

\lstdefinelanguage{json}{
    basicstyle=\normalfont\ttfamily,
    numbers=left,
    numberstyle=\scriptsize,
    stepnumber=1,
    numbersep=8pt,
    showstringspaces=false,
    breaklines=true,
    frame=lines,
    backgroundcolor=\color{background},
    literate=
     *{0}{{{\color{numb}0}}}{1}
      {1}{{{\color{numb}1}}}{1}
      {2}{{{\color{numb}2}}}{1}
      {3}{{{\color{numb}3}}}{1}
      {4}{{{\color{numb}4}}}{1}
      {5}{{{\color{numb}5}}}{1}
      {6}{{{\color{numb}6}}}{1}
      {7}{{{\color{numb}7}}}{1}
      {8}{{{\color{numb}8}}}{1}
      {9}{{{\color{numb}9}}}{1}
      {:}{{{\color{punct}{:}}}}{1}
      {,}{{{\color{punct}{,}}}}{1}
      {\{}{{{\color{delim}{\{}}}}{1}
      {\}}{{{\color{delim}{\}}}}}{1}
      {[}{{{\color{delim}{[}}}}{1}
      {]}{{{\color{delim}{]}}}}{1},
}
\newfloat{Code}{h}{myc}

\title{Engineering Sustainable Agents: A Systematic Comparison of Agentic LLMs for Developer Workflows}

\author{Merve Astekin}
\affiliation{%
  \institution{SINTEF
  \city{Oslo}
  \country{Norway}}
}
\email{merve.astekin@sintef.no}

\author{Yan Naing Tun}
\affiliation{%
  \institution{Singapore Management University \city{Singapore}
  \country{Singapore}}
}
\email{yannaingtun@smu.edu.sg}

\author{Arda Goknil}
\affiliation{%
  \institution{SINTEF
  \city{Oslo}
  \country{Norway}}
}
\email{arda.goknil@sintef.no}

\author{Erik Johannes Husom}
\affiliation{%
  \institution{SINTEF
  \city{Oslo}
  \country{Norway}}
}
\email{erik.johannes.husom@sintef.no}

\author{Lwin Khin Shar}
\affiliation{%
  \institution{Singapore Management University \city{Singapore}
  \country{Singapore}}
}
\email{lkshar@smu.edu.sg}

\author{Hasan Sözer}
\affiliation{%
  \institution{Ozyegin University
  \city{Istanbul}
  \country{Turkiye}}
}
\email{hasan.sozer@ozyegin.edu.tr}

\author{Ratnadira Widyasari}
\affiliation{%
  \institution{Singapore Management University \city{Singapore}
  \country{Singapore}}
}
\email{ratnadiraw.2020@phdcs.smu.edu.sg}

\author{Hui Song}
\affiliation{%
  \institution{SINTEF
  \city{Oslo}
  \country{Norway}}
}
\email{hui.song@sintef.no}

\renewcommand{\shortauthors}{Astekin et al.}

\begin{document}

\begin{abstract}


Large language models (LLMs) are increasingly used in software engineering, including agentic systems that coordinate multiple agents, but impose higher computational and environmental costs. In this paper, we present a comprehensive empirical study of agentic LLM systems across five software engineering tasks: code generation, technical debt identification, code vulnerability detection, log parsing, and log analysis. For each task, we compare LLM configurations that range from a non-agentic single-query baseline to multi-agent workflows, using six open-weight LLMs, two prompt strategies, and three 
hardware platforms. We assess each configuration in terms of accuracy, inference latency, and energy consumption. 
Our results reveal substantial trade-offs between agentic complexity and energy efficiency: multi-agent designs consume on average 6.36$\times$ as much energy and run 6.07$\times$ as long as the non-agentic baseline, with worst-case slowdowns of up to 160$\times$ for individual task--hardware pairs. Accuracy gains from additional agents are limited and task-specific: multi-agent improves average vulnerability-detection accuracy, but lightweight non-agentic and single-agent configurations still dominate the Pareto front, accounting for 59 of 66 Pareto-optimal configurations. Model and prompt choice act as task-specific levers whose effective direction varies between tasks rather than as global defaults. We translate these findings into 
design guidelines for sustainable, task-aware LLM-based development tools.

\end{abstract}

\maketitle

\section{Introduction}
\label{sec:introduction}

Large Language Models (LLMs) are transforming software engineering by supporting a variety of tasks, including code generation, debugging, code analysis, and documentation~\cite{ozkaya2023application, he2025llm, zhang2023survey, widyasari2025let, astekin2025detecting}. 
These capabilities have been accessed through single-query interactions with general-purpose models. On the other hand, the rise of LLM‑based multi‑agent systems, where multiple specialized agents collaborate to tackle different parts of a task, promises to enhance reasoning, problem decomposition, and robustness in developer workflows~\cite{he2025llm}. Early work in this area, such as agent frameworks for secure code generation and multi-agent debate mechanisms, suggests potential benefits including improved code correctness, enhanced security, and better scalability~\cite{nunez2024autosafecoder, ishibashi2024self, chen2025autoreview, pezze20252030, yu2025cxxcrafter}. At the same time, LLMs incur high environmental costs, including substantial energy consumption, water usage, and $\mathrm{CO_2}$ emissions, with impacts sometimes magnitudes greater for reasoning-intensive prompts compared to simpler ones~\cite{jegham2025hungry, husom2024price, samsi2023words}. This highlights an urgent need to consider sustainability alongside performance when designing agentic LLM systems.

\textbf{Problem Statement.} Despite the growing enthusiasm for LLM-based multi-agent systems in software engineering, the trade-offs between their effectiveness and sustainability remain under-investigated. Early work has demonstrated that agentic architectures, where autonomous agents orchestrate, refine, and reason collectively, can enhance task performance in domains like code generation and complex problem solving, with success rates increasing by up to 70\% compared to single-agent approaches~\cite{shu2024towards}. Similarly, literature reviews emphasize the transformative potential of such systems for autonomous software development workflows~\cite{he2025llm}. However, this enhanced capability often demands greater computational overhead. Previous studies have highlighted that \textbf{LLM inference consumes substantial energy and contributes notably to $\mathrm{CO_2}$ emissions}, especially for reasoning-intensive prompts~\cite{cheung2025comparative, samsi2023words, husom2024price, alizadeh2025language}.

Crucially, no existing study simultaneously evaluates agentic LLM configurations across multiple tasks, models, and hardware platforms with respect to both performance (accuracy, latency) and environmental cost (energy consumption). 
Without such a multi-dimensional, empirically grounded evaluation, practitioners lack the evidence needed to design effective yet sustainable multi-agent development tools. This gap calls for a structured analysis that bridges performance gains with ecological and practical deployment considerations.

\textbf{Goal of This Study.} The primary goal of this study is to systematically investigate the design, performance, and sustainability trade-offs of \textbf{multi-agent LLM systems} for automating software engineering tasks. Although multi-agent LLM architectures are increasingly studied across diverse domains, systematic empirical evidence on their effectiveness and efficiency in \textbf{practical developer tasks} remains scarce. This study aims to fill this gap by providing \textbf{empirical evidence and actionable} insights across a representative spectrum of software engineering tasks, including \textbf{code generation}, where LLMs are used to synthesize executable source code from natural language prompts; \textbf{technical debt identification}, which involves analyzing code snippets for signs of code smells; \textbf{code vulnerability detection,} which targets identifying security flaws or unsafe coding patterns; \textbf{log parsing,} where structured representations are extracted from raw log files; and \textbf{log analysis,} which involves interpreting parsed logs to detect errors or anomalous system behaviors.

We select these five software engineering tasks because they represent a diverse set of challenges across the software development lifecycle. They vary in terms of input/output structure, reasoning complexity, and reliance on external tools, making them well-suited for evaluating the performance and scalability of different LLM configurations. This diversity enables a comprehensive assessment of how agentic system design impacts effectiveness, efficiency, and sustainability in real-world development scenarios. For each task, we implement \textbf{a series of increasingly complex LLM-based configurations,} i.e., non-agentic single-query models, single-agent setups, dual-agent configurations, and multi-agent workflows involving collective intelligence and division of labor among agents.
These configurations are deployed using \textbf{multiple families of LLMs} (i.e., Gemma and its code-specialized variant CodeGemma, Qwen and its code-specialized variant Qwen-Coder, and DeepSeek-Coder) and evaluated on \textbf{different hardware platforms,} including a developer workstation and two high-performance servers.

Our goal is to evaluate each system setup along three critical dimensions: \textbf{accuracy}, to measure task-specific effectiveness; \textbf{inference latency}, to understand the responsiveness and feasibility of real-time deployment; and \textbf{energy consumption}, to assess the computational cost of running different configurations. 
We conduct this study with the expectation that trade-offs will emerge across multiple variation points, including the software engineering task, the underlying LLM model, the agentic system design, and the prompting strategy. This multi-dimensional evaluation enables us to \textbf{quantify the cost--benefit trade-offs} involved in deploying agentic LLM systems for software engineering and to derive \textbf{generalizable insights} into when and how such systems can be used effectively and sustainably.

\textbf{Research Questions.} We structure our study around four research questions (RQs), focusing on the effectiveness, efficiency, and sustainability of LLM-based agentic systems. Specifically, we investigate:

\begin{itemize}

\item How do different LLM system configurations perform across software engineering tasks
in terms of accuracy? (RQ1)

\item How do LLM system configurations differ in energy consumption 
across software engineering tasks and hardware platforms? (RQ2)

\item What is the inference latency of different agent configurations across software engineering
tasks and hardware platforms? (RQ3)

\item What trade-offs exist between performance, energy efficiency
, and latency across software engineering tasks and hardware platforms, and are there task-specific patterns that inform the design of sustainable multi-agent LLM systems? (RQ4)

\end{itemize}

\textbf{Our Findings.} Our results consistently show that agentic complexity has bounded benefits and systematic costs. The non-agentic baseline (NA) has the highest average accuracy for four of the five tasks, while the multi-agent configuration (MA) performs best for vulnerability detection. Moving from NA to MA increases energy consumption by 6.36$\times$ and latency by 6.07$\times$ on average, with worst-case slowdowns of up to 160$\times$. Of 66 Pareto-optimal configurations across all tasks, only one is multi-agent, and 36 (55\%) are non-agentic. Agentic configuration is therefore a first-order factor, but accuracy gains from additional agents are task-specific while the cost increase is much more systematic. Model and prompt choice, in contrast, have task-specific effects whose direction depends on the task: no single model wins across tasks, and prompt strategy alone can swing log-parsing accuracy from 3.4\% to 54.3\%. Hardware effects are modest for lightweight workloads but become first-order for heavier configurations, where a high-end server can be 3.4$\times$ faster than a workstation on MA. Energy and latency are tightly coupled (Pearson $r = 0.987$ in log--log space), so optimizing one effectively optimizes the other. Together, these findings indicate that competitive LLM-based developer tools can be built with lightweight, task-aware configurations, and that multi-agent designs should be adopted only with concrete evidence of a task-specific benefit.

\textbf{Our Contributions.} This paper makes the following key contributions:

\begin{itemize}

\item \textbf{A comprehensive multi-dimensional evaluation of LLM system configurations across diverse software engineering tasks.} We design and implement multiple LLM-based system configurations (ranging from non-agentic baselines to complex multi-agent workflows) for five representative software engineering tasks: code generation, technical debt identification, code vulnerability detection, log parsing, and log analysis. This task diversity allows us to assess how configuration complexity impacts effectiveness and efficiency in varied developer contexts.

\item \textbf{A systematic comparison of agentic configurations using different LLMs and hardware platforms.} We experiment with a range of instruction-tuned open-weight LLMs spanning multiple model families and sizes, and deploy them across heterogeneous hardware setups (i.e., workstations and servers). This enables us to investigate how model characteristics and deployment environments influence task performance and resource usage.

\item \textbf{A detailed empirical analysis of trade-offs between accuracy, latency, and energy consumption.} 
We measure each configuration's performance using a unified evaluation framework that captures inference latency and energy use. 
Our results reveal how agentic complexity, model size, prompting strategies, and hardware choice interact to affect performance and sustainability outcomes.

\item \textbf{Identification of task-specific and generalizable patterns across agentic design choices.} By comparing configurations across tasks, we uncover when lightweight agentic setups (e.g., single or dual agents) suffice and when more complex multi-agent workflows yield measurable benefits. We highlight configurations that offer favorable trade-offs, enabling developers to make informed design decisions based on their constraints and priorities.

\item \textbf{Design guidelines for sustainable and effective LLM-based developer tools.} Drawing from our findings, we distill a set of practical recommendations for building energy-efficient, task-aware, and scalable agentic LLM systems. These guidelines are intended to inform the future development of intelligent assistants, autonomous developer agents, and AI-augmented software engineering platforms.

\end{itemize}

Collectively, these contributions provide a comprehensive view of the effectiveness, efficiency, and sustainability of agentic LLM systems, offering guidance for their practical design and deployment in software engineering.

The remainder of this paper is organized as follows. In Section~\ref{sec:background}, we summarize the background on LLMs and energy monitoring. 
Section~\ref{sec:evaluation} presents our experimental setup, results, and analysis. In Section~\ref{sec:results}, we summarize our findings for each research question. Section~\ref{sec:discussion} synthesizes these findings across RQs and distils design guidelines for sustainable agentic LLM systems. Section~\ref{sec:validity} presents threats to the validity of our experiments. 
Finally, Section~\ref{sec:conclusion} concludes the paper and outlines directions for future research.

\section{Background and Related Work}
\label{sec:background}

\subsection{Multi-Agent LLM Systems}



Multi-agent systems (MAS) powered by Large Language Models (LLMs) are an emerging paradigm for enabling collaboration, specialization, and iterative reasoning across autonomous agents. In contrast to single-turn, single-agent interactions, multi-agent LLM systems leverage structured communication and role decomposition to handle complex tasks more effectively.

Recent surveys and systematizations outline the architectural components of LLM-based multi-agent systems, including agent roles, memory, planning, and interaction protocols. Li et al.~\cite{li2024survey} provide a comprehensive taxonomy of multi-agent LLM capabilities, highlighting key challenges in scalability, coordination, and self-evolution. 
Similarly, Tran et al.~\cite{TranDaoNguPhaSulNgu2025} and Liu et al.~\cite{LiuLoLuZhuZhaXuHarWhi2025} survey coordination strategies in multi-agent LLMs, offering categorizations of mechanisms and patterns that frame the design space of such systems.

Several open-source agent frameworks have been developed to support these patterns. They provide infrastructure for task orchestration, agent communication, and integration with external tools or data sources. Examples include \textbf{AutoGen}~\cite{wu2023autogen}, which enables multi-agent collaboration for solving complex tasks, and \textbf{Semantic Kernel}~\cite{semantickernel}, which combines LLMs with modular plugins for scalable AI systems. \textbf{AutoGPT}~\cite{autogpt} supports autonomous multi-step task execution through planning and refinement. Similarly, \textbf{OpenAgents}~\cite{xie2023openagents} and \textbf{AgentGPT}~\cite{agentgpt} focus on deploying autonomous AI agents, emphasizing modularity, integration, and usability for real-world applications. These frameworks excel in enabling multi-agent collaboration, where specialized agents—such as those for code generation, result interpretation, or output refinement—work together to achieve shared objectives. Frameworks like \textbf{AutoGen} and \textbf{AutoGPT} decompose complex workflows into manageable tasks, enhancing the efficiency and flexibility of LLM-powered systems. Interoperability is another critical feature, allowing seamless integration with external databases, APIs, and knowledge graphs. Tools like \textbf{OpenAgents}, \textbf{AgentGPT}, and \textbf{Semantic Kernel} streamline the processing of heterogeneous data, aligning real-time and historical information with system-specific knowledge to generate actionable insights.

Multi-agent LLM systems represent a significant evolution in AI-assisted software engineering, offering new capabilities for structured collaboration and task specialization. While frameworks such as AutoGen, OpenAgents, and AgentGPT provide a foundation for their development, rigorous empirical evaluation, particularly with respect to performance-efficiency trade-offs and environmental impact, is still lacking. This paper contributes to filling that gap by systematically assessing multi-agent configurations across software engineering tasks, models, and hardware platforms.

\subsection{Multi-Agent LLM Systems in Software Engineering}


Recent advancements in LLM-based MAS have sparked considerable interest in the software engineering community, owing to their enhanced problem-solving capabilities and structured collaboration patterns~\cite{he2025llm}. For instance, multi-agent frameworks built on LLMs have significantly advanced the field of automated code generation~\cite{jiang2024survey, he2025llm}. These approaches decompose the complex process of software synthesis into specialized sub-tasks, such as planning, coding, testing, and debugging, handled by collaborating agents, thereby addressing limitations in coherence, correctness, and scalability inherent to single-agent LLM systems. AgentCoder by Huang et al.~\cite{huang2023agentcoder} exemplifies this architecture with a system comprising programmer, test designer, and test executor agents. These agents iteratively generate, test, and refine code, resulting in performance gains on HumanEval and MBPP benchmarks over state-of-the-art single-agent models. Similarly, CodeTree~\cite{li2024codetree} introduces a tree-search-based planning strategy for LLM agents, which guides decision-making across code generation stages via both environment feedback and LLM-generated reasoning.

Another approach, CodeCoR, integrates self-reflective evaluation into the multi-agent code pipeline, combining prompt design, test generation, repair, and re-coding steps to ensure robustness and quality of results~\cite{pan2025codecor}. These frameworks often feature feedback loops where code is revised based on syntactic or semantic failures, a technique that has also been extended to security-enhanced frameworks such as AutoSafeCoder, which includes fuzz testing and static analysis agents to enforce safety constraints~\cite{nunez2024autosafecoder}. INTERVENOR~\cite{wang2023intervenor} employs a collaborative agent architecture consisting of a Code Learner and a Code Teacher. The Code Learner is responsible for generating an initial code implementation and compiling it to assess correctness. Upon detecting errors, the Code Teacher analyzes both the faulty code and the resulting error reports, then provides targeted repair instructions to guide the refinement of the solution. Self-repair~\cite{olausson2023self} and TGen~\cite{mathews2024test} improve code quality by leveraging feedback from the execution of predefined test cases to iteratively refine and correct generated programs. MAGIS~\cite{tao2024magis} introduces a four-agent system (Manager, Custodian, Developer, and QA) that cooperatively resolves GitHub issues, achieving up to an eightfold increase in resolution rates compared to GPT-4 alone. AgentMesh~\cite{khanzadeh2025agentmesh} structures agents into Planner, Coder, Debugger, and Reviewer roles that sequentially transform high-level requirements into tested, validated code, a practical orchestration of developer workflows. Self‑Organized Agents~\cite{ishibashi2024self} propose a dynamically scaled agent group to collaboratively generate and optimize ultra-large codebases. Their framework surpasses single-agent baselines by 5\% in pass@1 accuracy (i.e., the statistical probability that a single, independent generation attempt by an AI model correctly solves a given problem) on HumanEval through distributed, component-wise code generation.

While promising for software engineering tasks, multi-agent LLM systems also introduce new challenges in orchestration overhead, prompt alignment, tool integration, and reproducibility. Moreover, the environmental cost of multi-agent coordination remains largely unexplored. Increased agent interactions and tool calls often translate into higher inference latency and energy usage, raising important concerns around sustainability that existing frameworks do not address explicitly.


\subsection{Energy \& Sustainability in LLMs}


The environmental footprint of modern AI has become a central research concern. Early landmark analyses quantified the energy and carbon costs of training large NLP models and called for standardized reporting practices~\cite{strubell2019energy, henderson2020towards}. In parallel, the Green AI perspective argued for evaluating models not only by accuracy but also by efficiency metrics such as energy, emissions, and compute, to counter the escalating resource demands of state-of-the-art systems~\cite{schwartz2020green}. Subsequent systems-level work showed that lifecycle emissions depend strongly on model size, training/inference efficiency, datacenter efficiency (PUE), and the carbon intensity of the electricity mix, highlighting that location and infrastructure choices can change emissions by orders of magnitude~\cite{patterson2024energy}.

Beyond training, inference is now a dominant and fast-growing driver of AI’s footprint due to widespread deployment of LLMs. Studies emphasize that prompt length, decoding strategy, and multi-turn/agentic interaction patterns materially increase energy use and latency, raising the need for efficiency-aware evaluation and reporting during inference as well as training~\cite{cheung2025comparative, samsi2023words, husom2024price, alizadeh2025language, henderson2020towards, patterson2024energy, husom2025sustainable}. Tooling has emerged to help practitioners measure and report runtime impacts, e.g., CarbonTracker~\cite{anthony2020carbontracker} for training runs and epoch-level estimates, MLCO2/CodeCarbon-style approaches~\cite{lacoste2019quantifying} for standardized emissions estimation from energy data and regional grids, and MELODI~\cite{husom2024price} specialized for measuring energy consumption of LLM inference.

A complementary line of work targets algorithmic and systems optimizations to reduce inference cost: low-bit quantization and efficient kernels (e.g., GPTQ~\cite{frantar2022gptq} and LLM.int8~\cite{dettmers2022gpt3}), context and routing strategies that avoid unnecessary large-model calls (e.g., FrugalGPT~\cite{chen2023frugalgpt}), and faster decoding methods such as speculative decoding and multi-draft generation that cut token-generation latency and compute~\cite{leviathan2023fast, he2023rest}. 
Hardware-aware inference optimizations—including fused, I/O-aware attention kernels, paged KV-cache management, and KV-cache compression or eviction—reduce memory traffic and improve inference throughput on GPUs~\cite{dao2022flashattention,kwon2023efficient,zhang2023h2o}. However, their energy benefits depend strongly on the workload, software stack, and accelerator architecture~\cite{fernandez2025energy}. Hardware–software co-design for edge inference can further improve energy efficiency, reinforcing the need to jointly tune models, workloads, runtimes, and hardware for sustainable deployment~\cite{huang2025edgellm}.

Finally, sustainability extends beyond carbon and electricity. Resource accounting frameworks increasingly include water footprint and thermal constraints for datacenters, with evidence that large-scale LLM training and inference can incur substantial hidden water usage depending on cooling strategies and geography~\cite{li2025makingaithirstyuncovering}. Together, these strands motivate rigorous, multi-objective evaluation for accuracy and efficiency (energy, $\mathrm{CO_2}$, latency), and transparent reporting standards, especially for agentic LLM workflows where longer contexts and multi-turn exchanges can amplify resource consumption.

\subsection{Gaps in the Literature}



While the prior sections document advances in multi-agent LLM systems within software engineering and the growing attention to environmental implications of AI, three major gaps remain: 

\begin{itemize}

\item \textbf{Lack of Comprehensive Empirical Evaluations Across Agentic Configurations.} Most existing studies primarily evaluate their systems on isolated tasks (e.g., code generation or issue resolution), often comparing single-agent baselines to a single agentic design~\cite{he2025llm}. However, there is a lack of systematic empirical work that evaluates multiple agentic configurations (e.g., non-agentic, single-agent, dual-agent, multi-agent workflows) across a diverse set of software engineering tasks. This makes it difficult to generalize findings across development activities in the software lifecycle.

\item \textbf{Missing Joint Evaluation of Sustainability and Performance Metrics.} The growing corpus on Green AI and AI’s environmental impact emphasizes energy and $\mathrm{CO_2}$ metrics alongside accuracy (e.g.,~\cite{strubell2019energy, henderson2020towards, schwartz2020green, patterson2024energy, cheung2025comparative, samsi2023words, husom2024price}). However, no prior work in SE-focused multi-agent LLM systems has simultaneously measured accuracy, inference latency, energy consumption, and estimated carbon emissions for different architectural configurations. This omission hinders our ability to understand the trade-offs between performance and sustainability when deploying agentic systems.

\item \textbf{Reevaluating the Need for Agentic Complexity.} Recent work such as that by Xia et al.~\cite{xia2024agentless} questions whether elaborate agent architectures are always necessary. The authors show that a simpler, non-agentic three-phase approach (localization, repair, patch validation) can outperform complex agents in both performance and cost metrics on SWE‑bench Lite benchmarks. This suggests that agentic workflows may not always be the most efficient solution, and emphasizes the need for systematic comparison across configurations with sustainability in mind.

\end{itemize}

By explicitly addressing these gaps (multi-configuration comparison across tasks, integrated performance-sustainability measurement, and validation of agent simplicity), our study fills a critical void in both the SE and AI sustainability literature.

\section{Study Design}
\label{sec:evaluation}

This section presents the experimental setup (see Table~\ref{tab:experiment_setup} for a summary). We begin by outlining the research questions. Subsequently, we introduce the software engineering tasks and the datasets used for their evaluation. We then describe the agentic configurations examined in this study, followed by the implementation details and the models employed. Finally, we present the evaluation metrics, measurements, and the experimental protocol.

\subsection{Research Questions}
\label{subsec:research-questions}

We investigate the following Research Questions (RQs) across five software engineering tasks using four datasets, where two of the tasks (log parsing and log analysis) are performed on a shared dataset (HDFS logs).

\begin{itemize}

\item \textit{\textbf{RQ1.} How do different LLM system configurations perform across software engineering tasks in terms of accuracy?} This question aims to assess the effectiveness of each configuration, ranging from non-agentic to complex multi-agent workflows, in achieving accurate task completion across various software engineering scenarios.

\item \textit{\textbf{RQ2.} How do LLM system configurations differ in energy consumption 
across software engineering tasks and hardware platforms?} This question investigates the sustainability implications of different agent configurations by measuring their energy usage across deployment environments.

\item \textit{\textbf{RQ3.} What is the inference latency of different agent configurations across software engineering tasks and hardware platforms?} This question explores the responsiveness and real-time feasibility of different agent configurations, analyzing how architectural and hardware choices affect end-to-end inference time.

\item \textit{\textbf{RQ4.} What trade-offs exist between performance (accuracy), energy efficiency, and latency across software engineering tasks and hardware platforms, and are there task-specific patterns that inform the design of sustainable multi-agent LLM systems?} This question brings together the individual dimensions of RQ1--RQ3, exploring the multi-objective balance between performance, cost, and environmental impact, and deriving actionable design guidelines for practitioners.




\end{itemize}

\begin{table*}[t]
\centering
\caption{Overview of the empirical study design, including the software engineering tasks and datasets, agentic configurations, LLMs, hardware platforms, prompting strategies, measurement and execution protocols, statistical analyses, and their mapping to the four RQs. The experimental campaign comprises 1{,}800 runs from the core factorial design and 40 additional GPT-OSS runs on Server-SMU, for 1{,}840 runs in total.}
\label{tab:experiment_setup}
\scriptsize
\vspace*{-1.0em}

\begin{tabular}{p{2.8cm}p{10.2cm}}
\toprule
\textbf{Dimension} & \textbf{Experimental Setup} \\ 
\midrule

Software engineering tasks &
Code generation, technical debt identification, code vulnerability detection, log parsing, and log analysis. \\ \hline

Datasets &
HumanEval (code generation), MLCQ (technical debt detection), PrimeVul (code vulnerability detection), and LogHub's HDFS (log parsing and log analysis). \\ \hline

System configurations &
Non-agentic single-query models, single-agent setups, dual-agent configurations, and multi-agent workflows. \\ \hline

Hardware platforms &
Server-SMU (8× NVIDIA H100), Workstation-SMU (RTX A2000), and Server-STF (2× RTX 6000 Ada). \\ \hline

Models &
gemma3:4b, codegemma:7b-instruct, qwen3:4b-instruct, qwen2.5-coder:7b-instruct, deepseek-coder:6.7b-instruct, and gpt-oss:20b (Server-SMU only). \\ \hline

Prompting strategies &
Zero-shot and few-shot prompting (3 task-specific examples per prompt). \\ \hline

Total configurations &
5 tasks $\times$ 4 agent configurations $\times$ 2 prompting strategies $\times$ 5 models $\times$ 3 hardware platforms $\times$ 3 repetitions = $1800$ experimental runs, plus $40$ additional runs for gpt-oss:20b (evaluated once on Server-SMU), $1840$ in total. \\ \hline

Agent framework &
AG2 / AutoGen-style multi-agent orchestration framework. \\ \hline

Inference backend &
Ollama. \\ \hline

Energy measurement tool &
CodeCarbon with \texttt{tracking\_mode="process"} for process-level energy attribution. \\ \hline

Measurement levels &
CPU energy, GPU energy, and total pipeline energy (CPU + GPU + memory). \\ \hline

Carbon estimation &
CodeCarbon-based CO$_2$ estimation using a fixed Canadian carbon-intensity assumption. \\ \hline

Latency metric &
End-to-end inference latency measured per task instance. \\ \hline

Accuracy metrics &
Task-specific metrics including pass@k, precision, recall, F1-score, and parsing accuracy. \\ \hline

Execution protocol &
Temperature fixed at 0; Ollama restarted for each run; AG2 caching disabled; each experiment repeated three times where feasible, with GPT-OSS runs executed once due to resource constraints. \\ \hline

Statistical tests &
Friedman tests for multi-group paired comparisons, followed by Wilcoxon signed-rank post-hoc tests and matched-pairs rank-biserial effect sizes. \\ \hline

RQ1 design &
RQ1 compares accuracy across configurations, tasks, models, and hardware. \\ \hline

RQ2 design &
RQ2 compares process-level energy consumption across the same experimental dimensions; CO$_2$ emissions are derived from energy under a fixed carbon-intensity assumption. \\ \hline

RQ3 design &
RQ3 compares end-to-end inference latency across configurations, LLMs, hardware platforms, and tasks. \\ \hline

RQ4 design &
Multi-objective trade-off analysis comparing accuracy, energy consumption, and latency across configurations, models, tasks, and hardware platforms. \\ \hline

Reported results &
Configuration-level means and standard deviations across available repetitions, with aggregate tables averaging over the relevant configuration groups.\\ 
\bottomrule
\end{tabular}
\vspace*{-1.0em}

\end{table*}

\subsection{Tasks \& Datasets}
\label{subsec:datasets}

\subsubsection{Code Generation}
We used \textbf{HumanEval}, a benchmark introduced by Chen et al.~\cite{chen2021evaluating} for evaluating functional correctness in code generation with LLMs. 
The dataset consists of 164 handwritten Python programming problems, each including a function signature, natural language docstring, reference implementation, and several unit tests (on average 7.7 tests per problem)\footnote{\url{https://github.com/openai/human-eval}}. 
The problems are deliberately hand-written to avoid contamination from public sources such as GitHub repositories. 
HumanEval tasks cover language comprehension, reasoning, algorithmic design, and simple mathematics, making it a diverse and challenging benchmark that has been widely adopted for evaluating LLM-based code generation systems~\cite{huang2023agentcoder, ishibashi2024self, li2024codetree} 

\subsubsection{Technical Debt Detection}

We used an industry-relevant code smell dataset called \textbf{MLCQ} (Madeyski Lewowski Code Quest)~\cite{madeyski2020mlcq}. It contains nearly 15,000 code snippets in Java. Each snippet was independently reviewed by experts who were asked to indicate the presence or absence of a specific smell type. Two of these smell types are function-level (\emph{Long Method}, \emph{Feature Envy}). The other two are class-level (\emph{Blob}, \emph{Data Class}). Every code snippet that is checked for one of these smell types is labeled with one of the severity levels: \emph{None}, \emph{Minor}, \emph{Major}, \emph{Critical}. A single code snippet could be checked for multiple smells by the same or different experts. We analyzed recurring code snippets across different samples to consolidate a dataset for our evaluation. For the same code snippet with multiple reviews at either the function or class level:
\begin{itemize}
  \item If all reviews label the snippet with severity \emph{None}, we also assign it the \emph{None} label, indicating that it is not associated with any smell.
  \item If multiple smells are reported with severity other than \emph{None}, we exclude the snippet to avoid ambiguity, since the assigned task is to identify whether any smell was present and, if so, which one.
  \item If the severity for one review is \emph{None} and it is \emph{Major} or \emph{Critical} for the other type of smell at the corresponding level, we assign its label for the associated smell. For example, if the review for \emph{Long Method} indicates severity \emph{None}, but the review for \emph{Feature Envy} indicates severity \emph{Critical}, then we assign the label \emph{Feature Envy} with severity \emph{Critical} to that code snippet.
  \item We discarded entries labeled with \emph{Minor} severity, as well as those containing code snippets from inaccessible repositories, missing files, or outdated or hidden content.
\end{itemize}

As a result, we obtained a dataset with 6,441 samples without repeated entries (reviews) for any code snippet associated with one of the various types of smells. However, this dataset was still subject to code snippets that are possibly associated with multiple smell types at the same time. Reviewers assessing class-level code smells focused solely on class-level smell types~\cite{madeyski2020mlcq}. However, function-level smells can also exist within these classes. To avoid ambiguity and ensure that each code snippet is associated with a single smell type, we manually checked all the class-level code snippets. If a class-level code snippet contained a function-level smell, we excluded it from the dataset. For our evaluation, we selected a reduced sample of 385 code snippets as summarized in Table~\ref{tab:dataset}. These samples are randomly shuffled.

\begin{table}
  \centering
  \caption{Composition of the cleaned MLCQ evaluation subset by code-smell class and analysis granularity. Function-level samples include Long Method, Feature Envy, and smell-free functions; class-level samples include Blob, Data Class, and smell-free classes. ``None'' denotes samples without a target smell, and dashes indicate that a smell category is not applicable at the corresponding granularity. After duplicate removal and ambiguity filtering, the subset contains 385 samples in total.}
  \label{tab:dataset}
  \scriptsize
  \vspace*{-0.9em}

  \begin{tabular}{cccc}
    \toprule
    \textbf{Smell Type} & \textbf{Function Level} &  \textbf{Class Level} &  \textbf{Total} \\
    \midrule
    \emph{None} & 67 & 67 & 134 \\
    \emph{Blob} & - & 51 & 51 \\
    \emph{Data Class} & - & 67 & 67 \\
    \emph{Long Method} & 67 & - & 67 \\
    \emph{Feature Envy} & 66 & - & 66 \\
    \midrule
    \textbf{Total} & & & 385 \\
    \bottomrule
  \end{tabular}
  \vspace*{-0.9em}

\end{table}

\subsubsection{Code Vulnerability Detection}
We used \textbf{PrimeVul}, a benchmark for training and evaluating code language models (LMs) on vulnerability detection tasks~\cite{ding2023primevul}. 
Compared to other benchmarks, PrimeVul is designed to better capture the realistic challenges of detecting subtle vulnerabilities in source code. 
In this study, we specifically utilized the PrimeVul Pair subset, which presents a challenging benchmark as each vulnerable function is paired with a benign counterpart sharing at least 80\% of their string content. 
This design emphasizes fine-grained distinctions between secure and insecure implementations, making the task significantly more realistic and difficult. The dataset we utilize contains 193 vulnerable and 193 benign functions, resulting in a total sample size of 386.

\subsubsection{Log Parsing and Analysis}

For log-related tasks, we used datasets from \textbf{LogHub}, a large collection of system logs that are widely used for log parsing and anomaly detection research~\cite{zhu2023loghub}. 
LogHub contains over 77 GB of logs collected from both real-world production systems and controlled lab environments. 
The collection includes both labeled and unlabeled datasets: labeled logs are annotated for specific tasks such as anomaly detection or issue identification, while unlabeled logs are suitable for general-purpose log analysis.

In this study, we chose HDFS due to its widespread adoption in log analysis research.
The HDFS dataset contains logs from the Hadoop Distributed File System, a widely used distributed storage system. 
LogHub provides three HDFS variants (v1, v2, v3). 
We specifically used HDFS-v1, which was generated on a 203-node cluster running benchmark workloads, with anomalies manually labeled using handcrafted rules. 
The logs are organized into traces according to block IDs, and each trace is labeled as \emph{normal} or \emph{abnormal}. 
For our experiments, we used 385 log messages for the log parsing task and 385 sessions of log messages for the log analysis task, where each session corresponds to the logs associated with a specific block ID\footnote{\url{https://github.com/logpai/loghub}}.
To ensure that our experimental subset remained representative of the original dataset, we applied a stratified sampling strategy that preserves the original class distribution between normal and abnormal sessions for the log analysis task. In the HDFS dataset, anomalous sessions constitute approximately 3\% of the data; therefore, our sampled subset maintains this ratio. While this approach improves representativeness, it also results in a highly imbalanced dataset, which we take into account in our analysis.

\begin{table}[t]
\centering
\caption{Datasets and evaluation-set sizes for the five software engineering tasks. Sample counts denote HumanEval programming problems, cleaned MLCQ code snippets, PrimeVul Pair functions (193 vulnerable and 193 benign), HDFS-v1 log messages for log parsing, and HDFS-v1 block-ID sessions for log analysis. Although both HDFS tasks contain 385 samples, they use different units of analysis; the log-analysis subset was stratified to preserve the original normal-to-anomalous session distribution.}
\label{tab:dataset_stats}
\scriptsize
\vspace*{-0.9em}

\begin{tabular}{lrrr}
\toprule
\textbf{Task} & \textbf{Dataset} & \textbf{Samples} & \textbf{Source} \\
\midrule
Code Generation & HumanEval & 164 & \cite{chen2021evaluating} \\
Technical Debt & MLCQ & 385 & \cite{madeyski2020mlcq} \\
Vulnerability Detection & PrimeVul Pair & 386 & \cite{ding2023primevul} \\
Log Parsing & HDFS-v1 & 385 & \cite{zhu2023loghub} \\
Log Analysis & HDFS-v1 & 385 & \cite{zhu2023loghub} \\
\bottomrule
\end{tabular}
\vspace*{-0.9em}

\end{table}

\textbf{Rationale of the task and dataset selection:}
We selected the five software engineering tasks to cover complementary stages of the software lifecycle: code generation represents development-time automation, technical debt detection captures maintainability assessment, vulnerability detection addresses security analysis, while log parsing and log analysis represent operations and runtime system understanding. These tasks differ in reasoning complexity, input/output structure, and evaluation criteria, making them suitable for assessing whether agentic LLM configurations provide consistent benefits across heterogeneous software engineering scenarios.

For each task, we selected datasets that are widely used, publicly available, and suitable for reproducible evaluation. HumanEval was chosen for code generation because it provides executable programming problems with unit tests, enabling objective correctness measurement. MLCQ was selected for technical debt detection because it contains expert-reviewed code smell annotations at both function and class levels. PrimeVul was chosen for vulnerability detection because it provides realistic vulnerable and benign code pairs, supporting security-oriented code understanding. For log-related tasks, we selected LogHub/HDFS because it offers real-world log data with established templates and anomaly labels, enabling evaluation of both log parsing and log analysis under realistic operational conditions.


\subsection{System Configurations}
\label{subsec:configurations}

To systematically evaluate the impact of agentic collaboration on both task accuracy and energy consumption, we implemented four progressively collaborative configurations for each task, as shown in Figure~\ref{fig:configs}. Table~\ref{tab:config_summary} provides an overview of the configurations evaluated in this study.
Designing multi-agent LLM systems involves a large design space, including the number of agents, their roles, and their communication and coordination strategies. Consequently, there is no single universally optimal multi-agent architecture for all software engineering tasks. In this study, our objective is not to propose a new task-specific multi-agent design, but rather to evaluate the cost--benefit trade-offs associated with increasing levels of agentic collaboration. Figure~\ref{fig:configs} therefore illustrates a generic orchestration pattern. The concrete implementations used in our experiments adopt role configurations inspired by patterns previously reported in the literature, allowing us to evaluate representative and validated multi-agent setups rather than claiming task-optimal designs.
The set of configurations is explained in detail below:

\begin{figure}[t]
    \centering
    \includegraphics[width=0.95\linewidth]{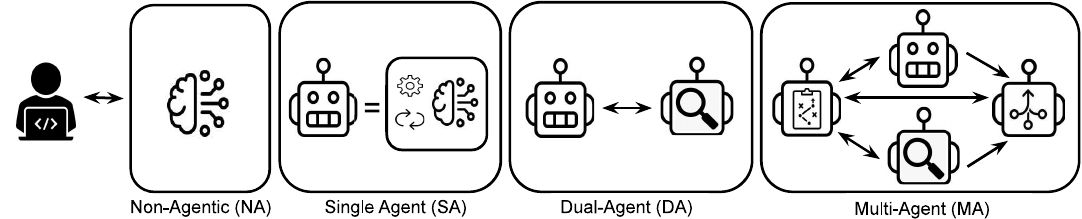}
    \caption{Overview of the four LLM system configurations evaluated in this study, ordered by increasing agentic complexity: non-agentic (NA), a direct stateless LLM query; single-agent (SA), one autonomous agent; dual-agent (DA), two collaborating task-specific agents; and multi-agent (MA), a coordinated four-agent workflow.}
    \label{fig:configs}
    \vspace*{-0.9em}

\end{figure} 

\begin{figure}[t]
    \centering
    \includegraphics[width=0.70\linewidth]{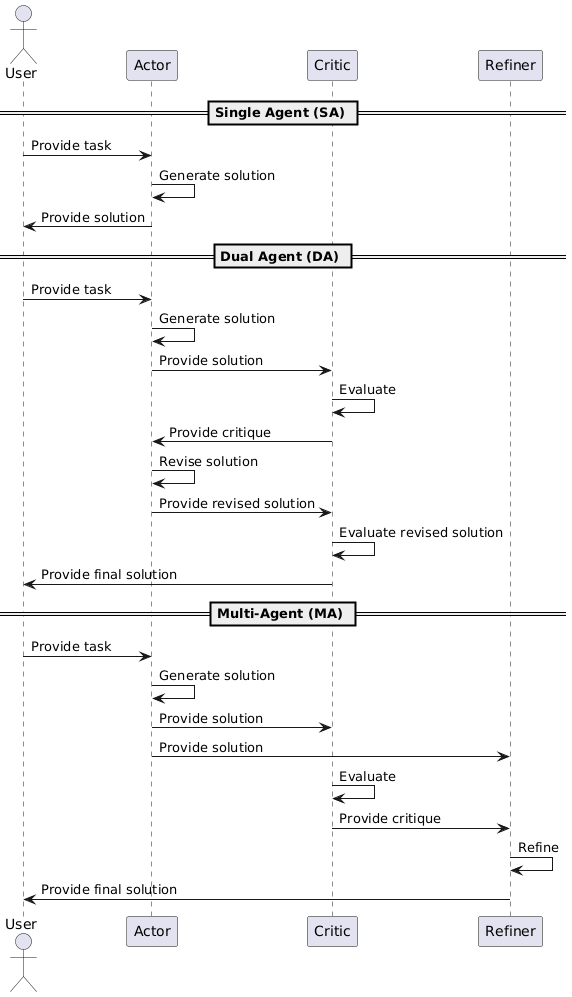}
    \vspace*{-0.7em}
    \caption{Sequence diagrams of the evaluated LLM system configurations. The non-agentic baseline issues a direct stateless query, whereas the single-agent configuration delegates the task to one autonomous agent. The dual-agent configuration adds a second agent for critique, refinement, or complementary analysis, and the multi-agent configuration coordinates four task-specific roles through a staged workflow that produces the final decision. Agent roles are instantiated separately for each software engineering task.}
    \label{fig:configScenarios}
    \vspace*{-1.0em}
\end{figure}

\begin{itemize}
    \item \emph{Non-Agentic (NA)}: This configuration involves direct calls to the LLM without maintaining any state or exhibiting agent-like behavior.
    \item \emph{Single Agent (SA)}: A single LLM-based agent acts autonomously to complete the task (See the top part of Figure~\ref{fig:configScenarios}).
    \item \emph{Dual Agent (DA)}: This configuration includes two LLM-based agents with task-specific roles. 
    For more structured tasks (\emph{log parsing and technical debt detection}), a general Actor--Critic setup is used, where the Actor produces an initial output, and the Critic evaluates and finalizes it.
    For tasks requiring deeper domain reasoning, specialized roles are defined: in \emph{code generation}, a Programmer who implements the solution and a Refiner who reviews and improves the Programmer's output to produce the final response; in \emph{vulnerability detection}, a Code Author who analyzes the code and explains its behavior, and a Security Analyst who reviews the Code Author's reasoning and the code itself to make the final vulnerability decision; in \emph{log analysis}, a Preprocessor parses the log messages in the given session and an Anomaly Detector that identifies whether the session contains anomalies based on the parsed output.
    \item \emph{Multi-Agent (MA)}: Four agents participate in a coordinated workflow with task-specific roles. The configurations follow established role-based cooperation, critique-and-refinement, and decision-aggregation patterns for LLM agents~\cite{LiuLoLuZhuZhaXuHarWhi2025}, and are implemented using the conversational orchestration mechanisms provided by AG2/AutoGen~\cite{wu2023autogen}. For structured tasks (\emph{log parsing} and \emph{technical debt detection}), the workflow consists of a User Proxy that provides the input, an Actor (Generator) that produces an initial output, a Critic that evaluates it, and a Refiner that incorporates the feedback into the final response. For domain-specific tasks, we define specialized roles according to the analytical steps required by each task. For \emph{code generation}, inspired by prior role-specialized generation and iterative review workflows~\cite{huang2023agentcoder}, a Requirements Analyst interprets the task, a Programmer implements the solution, a Moderator reviews its correctness and quality, and a Review Board makes the final assessment. For \emph{vulnerability detection}, we adopt the four-role mock-court configuration of VulTrial~\cite{widyasari2025let}: a Security Researcher identifies potential vulnerabilities, a Code Author responds to the findings, a Moderator summarizes the discussion, and a Review Board makes the final decision. For \emph{log analysis}, we introduce a task-specific decomposition in which a User Proxy provides the input session, a Preprocessor structures the log messages, an Anomaly Detector identifies potential anomalies, and a Critic reviews the findings and makes the final decision.

\end{itemize}

\begin{table}[t]
\centering
\scriptsize
\caption{LLM system configurations evaluated in the study, ordered by increasing agentic complexity. The ``\# Agents'' column counts active LLM-based agents, while the concrete DA and MA roles vary by software engineering task as described in Section~\ref{subsec:configurations}.}
\label{tab:config_summary}
\vspace*{-0.9em}

\begin{tabular}{lccl}
\toprule
\textbf{Configuration} & \textbf{\# Agents} & \textbf{Agentic} & \textbf{Key Characteristics} \\
\midrule
Non-Agentic (NA) & 0 & No & Direct LLM query, no state \\
Single Agent (SA) & 1 & Yes & Autonomous task completion \\
Dual Agent (DA) & 2 & Yes & Task-specific two-agent collaboration \\
Multi-Agent (MA) & 4 & Yes & Task-specific coordinated workflow \\
\bottomrule
\end{tabular}
\vspace*{-0.9em}

\end{table}

These configurations correspond to distinct collaboration mechanisms~\cite{TranDaoNguPhaSulNgu2025} and agent design patterns~\cite{LiuLoLuZhuZhaXuHarWhi2025}, each representing a different level of agentic interaction and orchestration. The \emph{NA} setup involves no collaboration or state maintenance, aligning with the \emph{one-shot model querying}~\cite{LiuLoLuZhuZhaXuHarWhi2025} pattern. It serves as a non-agentic baseline with minimal reasoning and energy cost, providing a lower bound for performance comparison. The \emph{SA} configuration corresponds to an agentic variant of the same pattern, where the LLM is encapsulated within an agent abstraction that manages prompts, execution flow, and internal state. Although \emph{NA} and \emph{SA} rely on a single LLM to complete the task and are therefore expected to yield comparable task performance, distinguishing them allows us to isolate and quantify the computational and energy overhead introduced by the agent abstraction itself. The \emph{DA} configuration introduces \emph{competitive collaboration}~\cite{TranDaoNguPhaSulNgu2025}, associated with the \emph{debate-based cooperation} and \emph{cross-reflection} patterns~\cite{LiuLoLuZhuZhaXuHarWhi2025}. This setup enhances reasoning consistency and output reliability through adversarial feedback, albeit with increased communication and energy demands. Finally, the \emph{MA} configuration embodies \emph{role-based coopetition}~\cite{TranDaoNguPhaSulNgu2025,LiuLoLuZhuZhaXuHarWhi2025} and \emph{voting-based cooperation}~\cite{LiuLoLuZhuZhaXuHarWhi2025} patterns, representing collective intelligence and division of labor among agents. This design is expected to produce more refined and robust task outcomes while introducing additional coordination overhead and energy consumption. 

\textbf{Rationale of the configuration selection:}
The four configurations were selected to span a representative and interpretable spectrum of agentic complexity, from a direct LLM call to a coordinated four-agent workflow, rather than to exhaustively cover all possible multi-agent designs. This progression allows us to isolate the marginal cost and benefit of each additional level of orchestration, agent abstraction overhead (NA-->SA), adversarial feedback (SA-->DA), and collective role-based reasoning (DA-->MA), while keeping the experimental space tractable. The specific role assignments within DA and MA were informed by agent design patterns reported in the literature~\cite{widyasari2025let, LiuLoLuZhuZhaXuHarWhi2025}, ensuring that the configurations reflect validated coordination strategies rather than arbitrary design choices. For structured tasks with well-defined input-output formats (log parsing, technical debt detection), generic Actor-Critic and User Proxy-Actor-Critic-Refiner patterns are appropriate because the task does not require domain-specific reasoning beyond what the prompt provides. For tasks involving nuanced code semantics or anomaly interpretation (code generation, vulnerability detection, log analysis), task-specific roles were assigned to give each agent a meaningful and distinct analytical function. 

\subsection{Implementation Details}
\label{subsec:implementation}
We evaluated the models using Ollama~\cite{MarcondesUsingOllama2025}, with each model configured using Ollama's default context length (4096).
To improve reproducibility and reduce output variability, all experiments were conducted with the temperature fixed at 0. 
Agent workflows were implemented using the AG2 framework (formerly AutoGen)\footnote{\url{https://github.com/ag2ai/ag2}}, an open-source platform for building multi-agent systems.

For prompt design, we adopted two strategies: zero-shot and few-shot prompting. 
In zero-shot prompting, the model receives only the task description, contextual information, and constraints, relying entirely on its pretrained knowledge. 
In few-shot prompting, the prompt additionally includes three manually selected labeled examples designed to capture diverse task-specific patterns (e.g., log event types, code smells, or code segments) to guide model behavior. These examples were not taken from the evaluation datasets to avoid data leakage. All prompt templates can be found in the replication package (\url{https://github.com/merveast/agent-green}). 




\subsection{Models}
\label{subsec:models}
In this study, we utilize five models from three model families (i.e., \emph{Gemma} family from Google, \emph{Qwen} family from Alibaba, and \emph{DeepSeek-Coder} family from DeepSeek AI) as listed below. 
Additionally, we evaluated \textbf{gpt-oss:20b} exclusively on one of the hardware configurations (the Server-SMU), as its 20B parameter size exceeded the capacity of the other hardware configurations.
\begin{itemize}
    \item \textbf{gemma3:4b}: A compact general-purpose instruction-tuned model from Google's Gemma familye~\cite{gemmateam2025gemma3technicalreport}. We include this model to represent lightweight, resource-efficient LLMs that can be deployed on constrained hardware while still supporting reasoning and code-related tasks. Gemma 3 models provide a large context window and are designed for efficient local deployment.
    
     \item \textbf{codegemma:7b-instruct:} A code-specialized instruction-tuned model from Google's CodeGemma family~\cite{codegemmateam2024codegemmaopencodemodels}. CodeGemma is designed specifically for programming-related tasks such as code generation, completion, and code understanding. We include it as a representative open-source code-oriented model for evaluating software engineering workflows.
     
    \item \textbf{qwen3:4b-instruct:} A compact instruction-tuned model from the Qwen3 family~\cite{yang2025qwen3technicalreport}. We include it to evaluate whether a smaller general-purpose model with strong instruction-following and reasoning capabilities can provide competitive results with lower energy and latency costs.
    
    \item \textbf{qwen2.5-coder:7b-instruct:} A code-specialized instruction-tuned model from the Qwen2.5-Coder family~\cite{hui2024qwen25codertechnicalreport}. It is optimized for code generation, code reasoning, and code repair, making it directly relevant to the software engineering tasks considered in this study.
    
     \item \textbf{deepseek-coder:6.7b-instruct:} A code-specialized instruction-tuned model trained on a mixture of code and natural language~\cite{guo2024deepseekcoderlargelanguagemodel}. We include it as another representative code-oriented model to compare against CodeGemma and Qwen2.5-Coder in terms of accuracy, latency, and energy efficiency. 
    
     \item \textbf{gpt-oss:20b (Server-SMU only):} A larger open-weight reasoning model from OpenAI's GPT-OSS family~\cite{openai2025gptoss120bgptoss20bmodel}. We include this model to assess how a substantially larger model affects the trade-offs between accuracy, inference latency, and energy consumption. Due to its parameter size and memory requirements, it was evaluated only on the Server-SMU configuration.
\end{itemize}

\textbf{Rationale of the model selection:} The selected models cover three complementary dimensions of the experimental design. First, they include both general-purpose instruction-tuned models and code-specialized models, allowing us to compare whether domain-specific code models provide measurable benefits for software engineering tasks. Second, they span different parameter scales, from compact 4B models to a larger 20B model, enabling us to analyze how model size affects accuracy, latency, energy consumption, and CO$_2$ emissions. Third, the models originate from different model families, which reduces the risk that the findings are specific to a single architecture or training pipeline. This selection, therefore, supports a balanced evaluation of model capability, deployment feasibility, and sustainability across heterogeneous LLM configurations.

\subsection{Hardware}
\label{subsec:hardware}


Table~\ref{tab:hardware} details the hardware platforms used in our data collection. The selected machines differ substantially in CPU architecture, memory capacity, GPU type, and available GPU memory, enabling us to evaluate how deployment environments influence the energy consumption, latency, and accuracy of LLM-based software engineering workflows.

\begin{table}[t]
    \scriptsize
    \caption{Hardware platforms used for experimental data collection, spanning two server-grade systems and one workstation-class system. The table reports the installed CPU, system memory, GPU configuration, and accelerator memory for each platform. GPU-memory values are given per accelerator, and the multiplicity indicates the number of installed GPUs rather than necessarily the number used concurrently by each experiment.}
    \label{tab:hardware}
    \centering
    \vspace*{-1.50em}

    \begin{tabular}{ccccc}
    \toprule
    \textbf{Machine} & \textbf{CPU} & \textbf{Memory (GB)} & \textbf{GPU} & \textbf{GPU memory (GB)} \\
    \midrule
    Server-SMU      & Intel Xeon Platinum 8480C & 2048 & 8 $\times$ NVIDIA H100 & 80 (HBM3) $\times 8$ \\
    Workstation-SMU  & Intel Xeon W-2223 8-core @ 3.60 GHz  & 128 & NVIDIA RTX A2000 & 12 \\
    Server-STF   & Intel Xeon w9-3475X 36-core @ 4.8GHz & 503 & 2 $\times$ NVIDIA RTX 6000 Ada & 48 $\times 2$ \\
    \bottomrule
    \end{tabular}
    \vspace*{-1.00em}
\end{table}

The \textit{Server-SMU} machine represents a high-performance server configuration with substantial memory and multiple high-end GPUs. This setup enables the execution of larger models and more complex agentic workflows that require higher memory capacity and parallel computation. In particular, this machine was used for configurations that could not be executed on the other hardware platforms due to model-size and GPU-memory constraints.

The \textit{Workstation-SMU} machine represents a mid-range workstation environment. It provides a more moderate CPU, memory, and GPU configuration, allowing us to evaluate how LLM-based workflows perform on hardware that is more accessible than large-scale server infrastructure. This setup is useful for studying whether lightweight or moderately complex agentic configurations can be executed efficiently without requiring high-end server resources.

The \textit{Server-STF} machine provides a second server-grade configuration with a different CPU generation and GPU architecture. By including this platform, we can compare the behavior of the same models and workflows across heterogeneous GPU environments and analyze whether energy consumption and latency trends remain consistent across different hardware setups.

\textbf{Rationale of the hardware selection:}
The hardware platforms were selected to cover a diverse range of realistic deployment settings for LLM-based software engineering tools. This diversity allows us to examine how hardware capabilities affect the feasibility and sustainability of different LLM configurations. The inclusion of both workstation- and server-grade machines enables comparison between accessible local deployment and more powerful infrastructure. Moreover, the variation in GPU memory and computational capacity allows us to study whether larger models and complex multi-agent workflows provide sufficient accuracy gains to justify their higher resource requirements. This hardware selection therefore supports a comprehensive analysis of performance--sustainability trade-offs across different computing environments.

\subsection{Energy Measurement}
\label{subsec:energy-metrics}
We used \textbf{CodeCarbon}~\cite{courty2024mlco2} as the tool for monitoring energy consumption during all experimental runs. 
CodeCarbon provides an end-to-end framework for collecting hardware-level power data and attributing it to a given process. 
It combines a power estimation model for RAM, Running Average Power Limit (RAPL) for CPU, and NVIDIA Management Library (NVML) for GPU. CodeCarbon was selected for its ease of integration and ability to provide reproducible, component-level energy estimates without requiring external sensors. All experiments were run on the hardware setups described previously. Table~\ref{tab:energy_measurement} gives an overview of our energy measurement setup.

CodeCarbon also estimates carbon emissions based on the energy consumption and carbon intensity of the energy grid in the geographic location where the experiments are run. We tracked carbon emissions during experiments using the offline mode of CodeCarbon, which is designed for environments without internet access. This mode preserves the same internal computation methodology but requires specifying a country\_iso\_code to approximate the carbon intensity of the electricity used. Since our computing resources were distributed across multiple geographic locations, we set the country code to CAN (Canada) as a representative average to ensure consistency across platforms. 
Canada's grid carbon intensity, as estimated by CodeCarbon, is lower than the global average and substantially lower than grids powered predominantly by fossil fuels; the CO$_2$ figures reported here should therefore be interpreted as conservative lower-bound estimates, and comparisons across studies should account for location-specific grid intensity. 
Since CO$_2$ emissions in this case are a linear, constant-factor transformation of energy consumption, it adds no information beyond the energy measurements and are left out of the results in Section~\ref{sec:results}.

\begin{table}[t]
\centering
\scriptsize


\caption{Energy-measurement configuration used in the experiments. CodeCarbon attributes CPU energy through Intel RAPL, GPU energy through NVIDIA NVML sampled at one-second intervals, and total pipeline energy as the combined CPU, GPU, and memory consumption recorded during each run. All values are reported in kWh using \texttt{tracking\_mode="process"} to estimate the energy attributable to the experimental process rather than the full machine.}
\vspace*{-0.90em}

\begin{tabular}{llp{5cm}p{1.7cm}}
\toprule
\textbf{Measurement Level} & \textbf{Data Source / Tool} & \textbf{What It Captures} & \textbf{Reporting Unit} \\ 
\midrule
CPU Energy & CodeCarbon (Intel RAPL) & Power draw of CPU package(s) during execution & kWh (aggregated per run) \\ 
GPU Energy & CodeCarbon (NVML interface) & Power draw of GPU(s) at 1-second intervals & kWh (aggregated per run) \\ 
Total Pipeline Energy & CodeCarbon (combined metrics) & Sum of CPU, GPU, and memory energy consumption over the run & kWh \\ 
\bottomrule
\end{tabular}
\label{tab:energy_measurement}
\vspace*{-0.9em}
\end{table}

We set CodeCarbon's tracking\_mode to "process," which estimates the power attributable specifically to the tracked process rather than the full-system ("machine") consumption. In this mode, CodeCarbon isolates the incremental resource usage (CPU via RAPL and GPU via pynvml) associated with the experiment, rather than reporting total node-level power draw.
As a result, the reported energy does not correspond to the full system’s idle or baseline power consumption (e.g., the substantial idle draw of an 8×H100 server). Instead, it reflects the estimated share of power usage attributable to the active processes under measurement. Consequently, no explicit subtraction of an idle baseline was performed, since the methodology inherently aims to exclude non-utilized resources from the accounting.
While this may underestimate total node-level energy, it provides a consistent basis for comparing experiments.

\subsubsection{Measurement Levels} Energy consumption was recorded at three levels:

\begin{itemize}

\item \textbf{CPU-level energy:} Captured using CodeCarbon’s integration with Intel RAPL counters to monitor CPU package power.

\item \textbf{GPU-level energy:} Measured through CodeCarbon’s NVIDIA NVML interface to track GPU power draw over time.

\item \textbf{Total pipeline energy:} Combined CPU, GPU, and memory measurements to reflect the overall energy used during model loading, inference, and agentic coordination.

\end{itemize}

By capturing both component-level and total energy use, we were able to analyze the relative contributions of CPU-bound and GPU-bound computation across different configurations.

We additionally recorded execution duration from CodeCarbon and used this value as the end-to-end inference latency metric in subsequent analyses.

%
%

\subsubsection{Reporting}
For each configuration-task pair, we ran three independent trials under identical conditions. Reported results include the mean energy consumption 
with standard deviation values to account for run-to-run variability. This ensures robust and reproducible sustainability measurements across all experimental scenarios.




\subsection{Latency \& Accuracy Measurement}
\label{subsec:latency-accuracy-metrics}

In addition to energy metrics, we evaluated each configuration on \textbf{latency} and \textbf{task-specific accuracy} to capture performance and user-perceived responsiveness. Together, these metrics provide a holistic view of the trade-offs between efficiency, quality, and sustainability.

\subsubsection{Latency Measurement}
Latency was measured as the end-to-end response time for each experimental run, from the moment a prompt was issued to the system until the final output was returned. For multi-agent configurations, latency included:

\begin{itemize}

\item \textbf{Model initialization time} (model loading from disk into memory),

\item \textbf{Agent orchestration overhead} (communication between agents, planning steps, and tool invocation),

\item \textbf{Model inference time} (token generation time for each agent involved).

\end{itemize}

We instrumented the execution pipeline to log \textbf{start and end timestamps} for each phase, allowing us to report both the total end-to-end latency and its internal breakdown (e.g., percentage of time spent in model loading vs. reasoning vs. tool use). Latency was measured in seconds, and results were averaged across three independent runs per configuration to mitigate transient fluctuations.

\begin{table}[t]
\centering
\caption{Task-specific effectiveness metrics and corresponding evaluation datasets. Higher values indicate better performance for all reported metrics.}
\label{tab:tasks_metrics}
\scriptsize
\vspace*{-1.2em}
\begin{tabular}{p{3.0cm}p{2.2cm}p{7.2cm}}
\toprule
\textbf{Task} & \textbf{Dataset} & \textbf{Accuracy Metric(s)} \\ 
\midrule

Code Generation &
HumanEval~\cite{chen2021evaluating} &
Pass@k \\ 

Technical Debt Identification &
MLCQ Dataset~\cite{madeyski2020mlcq} &
Accuracy, Macro F1, Per-Class F1 \\ 

Code Vulnerability Detection &
PrimeVul~\cite{ding2023primevul} &
Accuracy, Precision, Recall, F1-score \\ 

Log Parsing &
LogHub~\cite{zhu2023loghub} &
Parsing Accuracy (Exact Match), Edit Similarity, Longest Common Subsequence Similarity \\ 

Log Analysis &
LogHub~\cite{zhu2023loghub} &
Accuracy, Precision, Recall, F1-score \\ 
\bottomrule

\end{tabular}
\end{table}

\subsubsection{Accuracy Measurement}
Accuracy was assessed using \textbf{task-specific metrics} appropriate to each software engineering task. The five tasks produce different output structures (generated code, extracted templates, and class labels). Therefore, we use the term \emph{accuracy} as an umbrella term for task performance in the results, while reporting the metric that is most appropriate for each task. 
Table~\ref{tab:tasks_metrics} summarizes the tasks, datasets, and corresponding accuracy evaluation metrics used in the experiments.

\paragraph{Code Generation.}
For code generation, performance was measured using \textbf{Pass@k}, which estimates the probability that at least one of the top-$k$ generated solutions passes all unit tests~\cite{chen2021evaluating}.
In our experiments, we set $k=1$ since only a single solution is requested from the LLMs.

\paragraph{Log Parsing.}
Log parsing is evaluated as a structured prediction task, where the goal is to extract the correct log template from each raw log message.

We report three complementary metrics:

\begin{itemize}
    \item \textbf{Parsing Accuracy (Exact Match):} Measures the proportion of predicted templates that exactly match the ground-truth template, with no token deviations~\cite{he2016drain}.

    \item \textbf{Edit Similarity (ES):} A normalized edit-distance metric ranging from 0 to 1 that quantifies token-level similarity between predicted and ground-truth templates, where higher values indicate fewer edits~\cite{astekin2024acomparative}. ES captures partial correctness when templates are not exact matches.

    \item \textbf{Longest Common Subsequence (LCS) Similarity:} Measures normalized structural similarity based on the longest shared token subsequence between predicted and ground-truth templates, making it less sensitive to local token variations than ES~\cite{astekin2024acomparative}.
\end{itemize}

\paragraph{Classification Tasks.}
Technical debt identification, code vulnerability detection, and log analysis are treated as classification tasks.
For these tasks, performance is described using standard confusion matrix components:
True Positives (TP) are correctly identified positive instances, True Negatives (TN) are correctly identified negative instances, False Positives (FP) correspond to false alarms, and False Negatives (FN) represent missed positive instances.

Based on these quantities, we compute:

\begin{itemize}
    \item \textbf{Accuracy:} The proportion of correct predictions out of the total number of predictions.
    \[
    \text{Accuracy} = \frac{TP + TN}{TP + TN + FP + FN}
    \]

    \item \textbf{Precision:} The proportion of correctly identified positive instances out of all predicted positives.
    \[
    \text{Precision} = \frac{TP}{TP + FP}
    \]

    \item \textbf{Recall:} The proportion of correctly identified positive instances out of all actual positives.
    \[
    \text{Recall} = \frac{TP}{TP + FN}
    \]

    \item \textbf{F1-Score:} The harmonic mean of precision and recall, providing a balanced measure.
    \[
    \text{F1} = 2 \times \frac{\text{Precision} \times \text{Recall}}{\text{Precision} + \text{Recall}}
    \]
\end{itemize}

For \textbf{technical debt identification}, which is formulated as a multi-class code smell classification task with the classes \emph{No Smell}, \emph{Blob}, \emph{Data Class}, \emph{Feature Envy}, and \emph{Long Method}, we additionally report:

\begin{itemize}
\item \textbf{Macro F1:} Weights all classes equally and is therefore more informative than plain accuracy when comparing overall code smell detection capability across classes. 
    \[
    \text{F1}_{\text{macro}} = \frac{1}{|\mathcal{C}|}\sum_{c \in \mathcal{C}}\text{F1}_c
    \]
    where $\mathcal{C}$ is the set of technical debt classes. 
    
    \item \textbf{Per-Class F1:} One-vs-rest F1 scores for each code smell class, reported to identify which technical debt categories are comparatively easier or more challenging for each model to detect.
    
\end{itemize}

\subsection{Variables}
\label{subsec:variables}

In our experiment, we analyzed the \textbf{energy efficiency, inference performance, and output accuracy} of different LLM system configurations across multiple software engineering tasks, LLM models, and hardware platforms. These aspects are influenced by various factors, which can be categorized as \textbf{independent} and \textbf{dependent} variables.

\subsubsection{Independent Variables (Factors We Manipulate)}
\label{subsubsec:independent-variables}

These are the variables we control or modify to observe their effects on the dependent variables:

\begin{enumerate}

\item \textbf{Agent configuration} – The architectural setup of the LLM system, including  Non-Agentic (NA), Single-Agent (SA), Dual-Agent (DA), and Multi-Agent (MA) workflows.

\item \textbf{Software engineering task} – The specific task being performed: code generation, technical debt identification, code vulnerability detection, log parsing, or log analysis.

\item \textbf{LLM model} - The underlying large language model, varying across family, size, and specialization (general-purpose vs.\ code-oriented models).

\item \textbf{Hardware platform} – The execution environment: Server-SMU, Workstation-SMU, and Server-STF.

\item \textbf{Prompting strategy} – The method of interacting with the LLM (zero-shot or few-shot prompting).

\end{enumerate}

\subsubsection{Dependent Variables (Measured Outcomes)}
\label{subsubsec:dependent-procedure}

These are the variables we observe and measure to evaluate the impact of the independent variables:

\begin{enumerate}

\item \textbf{Energy Consumption} – Process-level energy consumed during execution, measured in Joules or kWh.



\item \textbf{Inference latency} – End-to-end execution time from task submission to final response generation, including model inference and agent coordination overhead.

\item \textbf{Output Accuracy} – Task-specific correctness metrics including Pass@1, parsing accuracy, edit similarity, LCS similarity, accuracy, precision, recall, F1-score, Macro F1, and per-class F1, depending on the task.


\end{enumerate}

By analyzing the impact of independent variables, we aimed to gain insights into how different design choices in LLM-based system configurations, task types, model selection, hardware platforms, and prompting strategies influence the trade-offs between accuracy, efficiency, and sustainability in software engineering applications.

\subsection{Experiment Execution}
\label{subsec:experiment-execution}

\begin{figure}[t]
\centering
\includegraphics[width=1.0\linewidth]{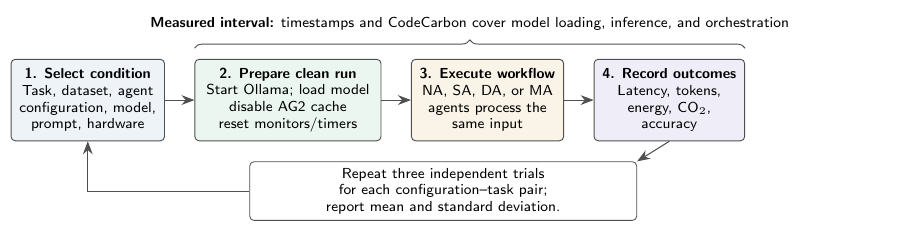}
\vspace*{-1.40em}
\caption{Overview of experimental workflow showing model initialization, execution, and measurement phases}
\vspace*{-0.90em}
\label{fig:experimental_workflow}
\end{figure}

To ensure that our measurements of energy consumption, latency, and accuracy were both consistent and reproducible, we implemented a strict and repeatable experiment execution protocol. Each experimental trial followed the same initialization, execution, and termination sequence to minimize uncontrolled variability and prevent cross-run interference. Figure~\ref{fig:experimental_workflow} illustrates the overall experimental workflow.

\begin{itemize}

\item \textbf{Model Initialization Control:} For each experimental run, the Ollama server was explicitly started immediately before execution and terminated immediately afterward. This ensured that each run began with a cold start, preventing the reuse of a model already loaded into memory. This step is essential because keeping a model resident in memory can artificially reduce both loading time and initial energy draw, thus biasing the measurements. By forcing the model to be loaded fresh for every run, we accurately captured the true energy cost and latency associated with model initialization.
Additionally, Ollama's \textit{keep\_alive} parameter was set to $-1$ to keep the model loaded in memory throughout the duration of each run. This prevented mid-run model eviction between the sequential LLM calls, avoiding spurious latency and energy overhead from unintended reloads.

\item \textbf{Elimination of Agent-Level Caching:} Within the AG2 multi-agent framework, all forms of agent-level caching were disabled. This prevented agents from retrieving previously stored intermediate results or responses, ensuring that each prompt and sub-task was processed from scratch. Disabling caching eliminates potential biases in accuracy measurement (since cached answers might artificially inflate success rates) and ensures that computation and communication energy costs reflect actual agent reasoning and tool usage.

\item \textbf{Isolated Measurement Environment:} Each run was executed in an isolated environment without concurrent workloads. This minimized background resource contention, such as CPU scheduling delays or GPU utilization spikes, that could otherwise affect latency and energy measurements. System resource monitoring tools were reset prior to each run to prevent accumulation of previous statistics.

\item \textbf{Repetition for Statistical Robustness:} To account for run-to-run variability (e.g., minor fluctuations in system performance, hardware temperature, or network conditions), each configuration–task pair was executed three independent times. The reported results are the arithmetic mean of these repetitions. In addition, standard deviation values were calculated to assess stability and reliability.

GPT-OSS configurations were executed with a single repetition only due to substantially higher computational requirements.

\item \textbf{Consistent Input and Execution Parameters:} All prompts, datasets, and execution parameters were kept identical across repetitions for a given experimental condition. The temperature was fixed at 0 for all experiments to reduce output variability. Random seeds were fixed where applicable (e.g., in stochastic decoding strategies such as sampling) to reduce nondeterministic effects while still preserving the typical behavior of the LLM.

\end{itemize}

By adhering to this controlled procedure, we ensured that observed differences in energy consumption, latency, or accuracy 
could be confidently attributed to changes in the independent variables (e.g., system configuration, LLM model, hardware platform) rather than to measurement noise or uncontrolled environmental factors.

\subsection{Data Analysis}
\label{subsec:data-analysis}

\textbf{Data Preparation.} Our raw data includes energy consumption, processing time, and task-specific accuracy metrics per inference operation. The data is recorded in spreadsheet format, organized by software engineering task, with one sheet per task. Each row corresponds to one experimental run and records the configuration parameters (agent design, LLM model, hardware platform, prompting strategy, and repetition round) alongside the measured outcomes: energy consumption in kWh, compute time in seconds, and the task-specific accuracy metric. The full experiment produced a total of 1{,}840 individual runs across all five tasks, resulting from the cross-product of the independent variables with three repetitions per configuration (except for GPT-OSS configurations, which were run once due to resource constraints).


\textbf{Data Post-processing.} 
The first step 
involves normalizing and cleaning the raw records. Column values for agent design, model name, hardware platform, and prompt type are standardized to canonical forms (e.g., consistent casing, whitespace trimming, and removal of timestamps from model identifiers). The cleaned run-level data is then aggregated into configuration-level summaries by grouping on the five configuration dimensions (task, agent, model, hardware, prompt) and computing the mean and standard deviation of each dependent variable across the repeated runs. This yields 640 unique configuration-level records, each summarizing up to three replicate runs.

\textbf{Visualization.} We employed three visualization techniques to represent our analysis:
\begin{enumerate}
    \item Box plots: To visualize the distribution of energy consumption metrics across different models and agentic configurations. Each box plot depicts the median (horizontal line splitting the main box), interquartile range (IQR, main box), and whiskers extending to 1.5 times the IQR, effectively summarizing central tendency and variability.
    \item Scatter plots: To examine relationships between energy consumption and accuracy, with points representing different model configurations. Data points are colored and styled according to which model and dataset they represent. For certain figures, we plot the fitted regression lines, indicating trend lines.
    \item Pareto frontier curves: Connecting Pareto-optimal configurations to visualize the trade-off boundary between energy efficiency and accuracy.
\end{enumerate}





\subsection{Replication Package}
\label{sec:replication-package}

To support independent replication and extension of this study, we provide a replication package at \url{https://github.com/merveast/agent-green}. The package contains the source code used to implement the non-agentic, single-agent, dual-agent, and multi-agent configurations; the task-specific agent roles and interaction workflows; and the zero-shot and few-shot prompt templates. It also includes the scripts used to prepare the evaluation subsets from HumanEval, MLCQ, PrimeVul Pair, and HDFS-v1, together with the sampled instance identifiers and preprocessing rules needed to reconstruct the experimental datasets.

The package further provides model and Ollama configurations, AG2 workflow definitions, CodeCarbon instrumentation, hardware and software environment specifications, and scripts for executing the experiments and collecting task outputs, latency, and component-level energy measurements. 
Aggregated per-configuration results are provided in a structured spreadsheet, together with the scripts used to calculate task-specific accuracy metrics, descriptive statistics, statistical tests, effect sizes, and Pareto fronts are included to enable verification of the reported findings. For models or datasets that cannot be redistributed directly because of licensing or size restrictions, the package provides acquisition instructions, version identifiers, and preprocessing scripts rather than copies of the original artifacts.

A detailed README documents environment setup, dependency installation, experiment execution, expected directory structure, and result reproduction. Machine-readable configuration files record the model, prompt strategy, agent configuration, task, hardware platform, and repetition associated with each run. Because absolute energy and latency measurements depend on the deployment environment, exact numerical reproduction may require comparable hardware; nevertheless, the supplied artifacts support replication of the experimental procedure and evaluation pipeline on alternative platforms.

\section{Experiment and Results}
\label{sec:results}

This section discusses the results of our case studies, addressing each RQ in turn. Each subsection reports the effect of the four main factors (agent configuration, hardware platform, prompt strategy, and LLM model) in the same order, followed where applicable by task-specific details.

\paragraph{Scope of cross-hardware aggregates.} The GPT-OSS model was evaluated on a single hardware platform (Server-SMU) and without repeated runs due to its 20-billion-parameter footprint. To avoid biasing cross-hardware comparisons, all hardware-balanced aggregates exclude GPT-OSS unless explicitly noted. Model-level aggregates that span all hardware platforms include GPT-OSS only on Server-SMU; we flag the resulting interpretation caveats where they apply.

\subsection{Effectiveness of Configurations (RQ1)}
\label{subsec:RQ1}

RQ1 investigates how the LLM system configurations perform across the five software engineering tasks, by reporting task-specific performance metrics for each task: Pass@1 for code generation, Macro F1 for technical debt detection, F1-score for vulnerability detection and log analysis, and parsing accuracy for log parsing.

\subsubsection{Effect of agentic configuration.}
\begin{figure}
    \centering
    \includegraphics[width=1.0\linewidth]{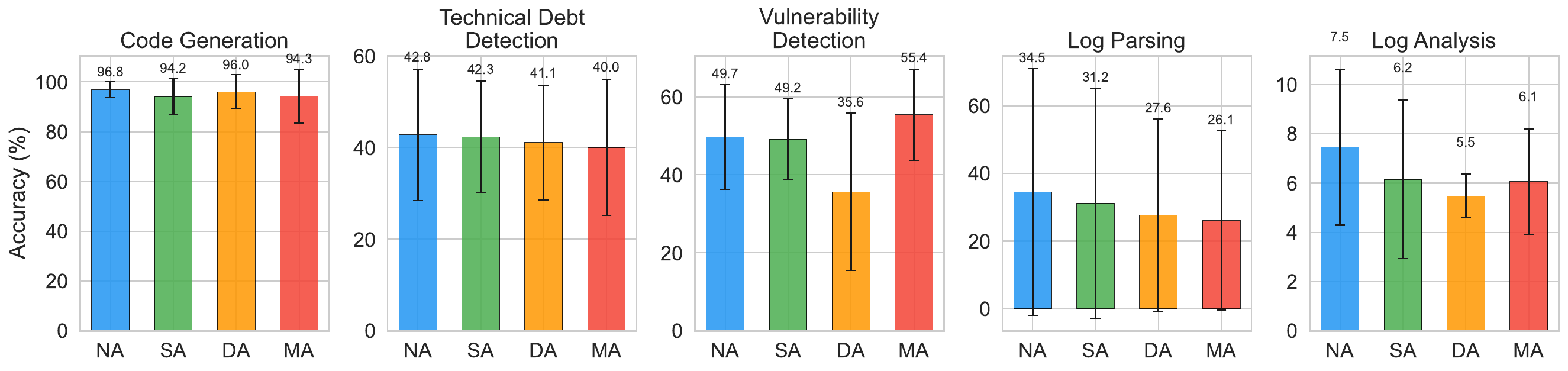}
    \vspace*{-0.90em}
    \caption{RQ1 task-specific effectiveness across LLM system configurations. Bars and error bars report the mean $\pm$ one standard deviation across model--prompt combinations after first averaging over the available hardware platforms within each combination. Performance is measured using Pass@1 for code generation, Macro F1 for technical debt detection, F1-score for vulnerability detection and log analysis, and exact-match parsing accuracy for log parsing. NA denotes the non-agentic direct-query baseline, SA the single-agent configuration, DA the dual-agent configuration, and MA the four-agent multi-agent configuration.}
    \label{fig:rq1_accuracy}
\end{figure}

Figure~\ref{fig:rq1_accuracy} summarizes the average performance of the four configurations. The main pattern is that increasing agentic complexity does not systematically improve accuracy, but the vulnerability-detection results provide an important exception. The non-agentic baseline (NA) obtains the highest mean score for code generation, technical debt detection, log parsing, and log analysis, whereas MA obtains the highest mean score for vulnerability detection.

For code generation, all configurations perform strongly, with Pass@1 ranging from 94.2\% for SA to 96.8\% for NA. This suggests that, for the HumanEval-style tasks considered here, additional agentic orchestration is not necessary to achieve high functional correctness. Technical debt detection shows a similarly flat pattern across configurations, with Macro F1 between 40.0\% and 42.8\%. The low absolute scores indicate that the task is challenging, but the differences between agentic designs are small.

Vulnerability detection is the clearest exception to the overall pattern. MA obtains the highest average F1-score (55.4\%), followed by NA (49.7\%) and SA (49.2\%), while DA remains substantially lower (35.6\%). Thus, the multi-agent workflow appears beneficial for this task, but the benefit is specific to MA rather than to agentic complexity in general.

For the two log tasks, the results are weaker overall and more variable. Log parsing reaches 34.5\% parsing accuracy for NA, followed by 31.2\% for SA, 27.6\% for DA, and 26.1\% for MA. Log analysis has low F1-scores for all configurations, ranging from 5.5\% to 7.5\%. These results indicate that the log tasks remain difficult for the evaluated models and prompts, and that adding agents does not by itself solve the underlying extraction and classification challenges.
For log analysis in particular, the uniformly low F1-scores are primarily attributable to severe class imbalance: anomalous sessions constitute approximately 3\% of the HDFS dataset, and our experimental subset preserves this ratio by design to maintain representativeness. The reported F1-scores therefore reflect the difficulty of detecting rare anomalies rather than a general failure of the configurations, and absolute scores for log analysis should not be compared directly with those of the other tasks.

\paragraph{Statistical analysis.}
Table~\ref{tab:significant_factors_all_tasks} shows which design factors show statistically significant effects across tasks for both energy consumption and task performance, using Friedman and Wilcoxon signed-rank tests. To align the statistical tests with the RQ1 accuracy aggregation, repeated runs are first averaged within each configuration, and hardware is then averaged within each task--agent--model--prompt combination. Thus, each observation in the accuracy test corresponds to one paired, hardware-balanced model--prompt cell rather than to an individual repeated run. Under this aggregation, agentic configuration has a statistically significant effect on accuracy only for vulnerability detection ($p=0.0162$). The effect is not significant for code generation ($p=0.830$), technical debt detection ($p=0.0752$), log parsing ($p=0.666$), or log analysis ($p=0.284$). By contrast, energy consumption differs significantly by agentic configuration for all tasks.

The vulnerability-specific post-hoc tests in Table~\ref{tab:stest-accuracy-vul-detection} show that DA differs significantly from MA, NA, and SA on accuracy, while the differences among MA, NA, and SA are not significant under the paired hardware-balanced test. The corresponding energy comparison in Table~\ref{tab:stest-energy-vul-detection} shows significant differences between DA and MA and between MA and NA for vulnerability detection. 

Overall, the statistical analysis in Table~\ref{tab:significant_factors_all_tasks} supports the pattern that additional agents do not provide a general accuracy benefit, but they do systematically affect resource consumption.
The choice of model shows significant difference on energy consumption across all tasks, while it only shows significance for the accuracy on code generation, technical debt detection and vulnerability detection.
Prompt type is not consistently significant across the tasks for any metric. Hardware shows statistical significance for both accuracy and energy consumption across all tasks, except the accuracy of vulnerability detection and log analysis.

\begin{table}[t]
\centering
\caption{RQ1 pairwise Wilcoxon signed-rank comparisons among the non-agentic (NA), single-agent (SA), dual-agent (DA), and four-agent multi-agent (MA) configurations for vulnerability-detection F1-score, following the significant omnibus Friedman test. Repeated runs are first averaged within each configuration, and hardware is then averaged within each task--agent--model--prompt combination, yielding paired hardware-balanced model--prompt observations. Statistical significance is assessed at $p<0.05$. Matched-pairs rank-biserial correlation ($r$) reports effect size and direction: positive values favor Configuration~1, negative values favor Configuration~2, and larger absolute values indicate stronger differences.}
\label{tab:stest-accuracy-vul-detection}
\scriptsize
\begin{tabular}{lllll}
\hline
Configuration 1 & Configuration 2 & $p$-value & Significant & Rank-biserial $r$ \\
\hline
DA & MA & 0.0122 & Yes & -0.795 \\
DA & NA & 0.0122 & Yes & -0.795 \\
DA & SA & 0.0161 & Yes & -0.769 \\
MA & NA & 0.339 & No & +0.333 \\
MA & SA & 0.339 & No & +0.333 \\
NA & SA & 0.677 & No & +0.154 \\
\hline
\end{tabular}
\end{table}

\begin{table}[t]
\centering
\caption{RQ2 pairwise Wilcoxon signed-rank comparisons of total pipeline energy consumption among the non-agentic (NA), single-agent (SA), dual-agent (DA), and four-agent multi-agent (MA) configurations for vulnerability detection, following the significant omnibus Friedman test. Repeated runs are first averaged within each configuration, and hardware is then averaged within each agent--model--prompt combination, yielding paired hardware-balanced model--prompt observations. Statistical significance is assessed at $p<0.05$. Matched-pairs rank-biserial correlation ($r$) reports effect size and direction: positive values indicate higher energy consumption for Configuration~1, negative values indicate higher consumption for Configuration~2, and larger absolute values indicate stronger differences.}
\label{tab:stest-energy-vul-detection}
\scriptsize
\begin{tabular}{lllll}
\hline
Configuration 1 & Configuration 2 & $p$-value & Significant & Rank-biserial $r$ \\
\hline
DA & MA & 0.0049 & Yes & -0.872 \\
DA & NA & 0.850 & No & +0.077 \\
DA & SA & 0.910 & No & +0.051 \\
MA & NA & 0.0210 & Yes & +0.744 \\
MA & SA & 0.0771 & No & +0.590 \\
NA & SA & 0.424 & No & +0.282 \\
\hline
\end{tabular}
\end{table}

\begin{table*}[t]
\centering
\caption{Factors showing paired-test differences across tasks. Multi-level factors use Friedman tests, while two-level prompt comparisons use Wilcoxon signed-rank tests. Agent, model, and prompt tests use hardware-balanced observations; hardware tests use configuration means because hardware cannot be tested after it has been averaged out. Significant differences ($p < 0.05$) are highlighted in bold.}
\label{tab:significant_factors_all_tasks}
\scriptsize
\begin{tabular}{llllr}
\toprule
Task & Metric & Factor & Test & $p$-value \\
\midrule
\multirow{8}{*}{Code Generation} & Pass@1 & Agent Design & Friedman & 0.830 \\
 & \ubf{Pass@1} & \ubf{Model} & \ubf{Friedman} & \ubf{0.0018} \\
 & \ubf{Pass@1} & \ubf{Prompt Type} & \ubf{Wilcoxon} & \ubf{$8.65 \times 10^{-4}$} \\
 & \ubf{Pass@1} & \ubf{Hardware} & \ubf{Friedman} & \ubf{0.0387} \\
 & \ubf{Energy} & \ubf{Agent Design} & \ubf{Friedman} & \ubf{$3.50 \times 10^{-5}$} \\
 & \ubf{Energy} & \ubf{Model} & \ubf{Friedman} & \ubf{0.0077} \\
 & Energy & Prompt Type & Wilcoxon & 0.345 \\
 & \ubf{Energy} & \ubf{Hardware} & \ubf{Friedman} & \ubf{$3.52 \times 10^{-13}$} \\
\midrule
\multirow{8}{*}{Technical Debt Detection} & Macro F1 & Agent Design & Friedman & 0.0752 \\
 & \ubf{Macro F1} & \ubf{Model} & \ubf{Friedman} & \ubf{$4.03 \times 10^{-7}$} \\
 & Macro F1 & Prompt Type & Wilcoxon & 0.603 \\
 & \ubf{Macro F1} & \ubf{Hardware} & \ubf{Friedman} & \ubf{0.0360} \\
 & \ubf{Energy} & \ubf{Agent Design} & \ubf{Friedman} & \ubf{$1.98 \times 10^{-7}$} \\
 & \ubf{Energy} & \ubf{Model} & \ubf{Friedman} & \ubf{$3.64 \times 10^{-6}$} \\
 & \ubf{Energy} & \ubf{Prompt Type} & \ubf{Wilcoxon} & \ubf{0.0072} \\
 & \ubf{Energy} & \ubf{Hardware} & \ubf{Friedman} & \ubf{$3.77 \times 10^{-10}$} \\
\midrule
\multirow{8}{*}{Vulnerability Detection} & \ubf{F1-score} & \ubf{Agent Design} & \ubf{Friedman} & \ubf{0.0162} \\
 & \ubf{F1-score} & \ubf{Model} & \ubf{Friedman} & \ubf{0.0255} \\
 & F1-score & Prompt Type & Wilcoxon & 0.565 \\
 & F1-score & Hardware & Friedman & 0.509 \\
 & \ubf{Energy} & \ubf{Agent Design} & \ubf{Friedman} & \ubf{0.0213} \\
 & \ubf{Energy} & \ubf{Model} & \ubf{Friedman} & \ubf{$1.97 \times 10^{-5}$} \\
 & Energy & Prompt Type & Wilcoxon & 0.0894 \\
 & \ubf{Energy} & \ubf{Hardware} & \ubf{Friedman} & \ubf{$7.92 \times 10^{-17}$} \\
\midrule
\multirow{8}{*}{Log Parsing} & Parsing accuracy & Agent Design & Friedman & 0.666 \\
 & Parsing accuracy & Model & Friedman & 0.0578 \\
 & \ubf{Parsing accuracy} & \ubf{Prompt Type} & \ubf{Wilcoxon} & \ubf{$3.58 \times 10^{-7}$} \\
 & \ubf{Parsing accuracy} & \ubf{Hardware} & \ubf{Friedman} & \ubf{0.0124} \\
 & \ubf{Energy} & \ubf{Agent Design} & \ubf{Friedman} & \ubf{$5.49 \times 10^{-7}$} \\
 & \ubf{Energy} & \ubf{Model} & \ubf{Friedman} & \ubf{$1.31 \times 10^{-4}$} \\
 & Energy & Prompt Type & Wilcoxon & 0.264 \\
 & \ubf{Energy} & \ubf{Hardware} & \ubf{Friedman} & \ubf{$4.52 \times 10^{-13}$} \\
\midrule
\multirow{8}{*}{Log Analysis} & F1-score & Agent Design & Friedman & 0.284 \\
 & F1-score & Model & Friedman & 0.243 \\
 & F1-score & Prompt Type & Wilcoxon & 0.277 \\
 & F1-score & Hardware & Friedman & 0.251 \\
 & \ubf{Energy} & \ubf{Agent Design} & \ubf{Friedman} & \ubf{$7.35 \times 10^{-7}$} \\
 & \ubf{Energy} & \ubf{Model} & \ubf{Friedman} & \ubf{$2.10 \times 10^{-4}$} \\
 & \ubf{Energy} & \ubf{Prompt Type} & \ubf{Wilcoxon} & \ubf{0.0039} \\
 & \ubf{Energy} & \ubf{Hardware} & \ubf{Friedman} & \ubf{$1.01 \times 10^{-13}$} \\
\bottomrule
\end{tabular}
\end{table*}

\subsubsection{Effect of hardware platform.}
\begin{figure}
    \centering
    \includegraphics[width=0.72\linewidth]{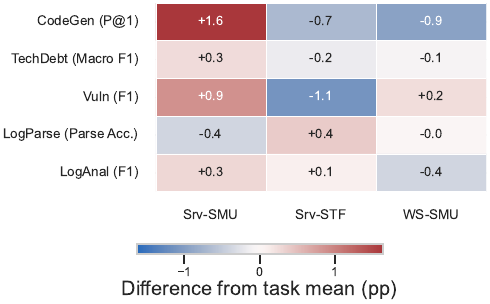}
    \vspace*{-0.90em}
    \caption{RQ1 hardware effects on task-specific performance using the hardware-balanced subset of configurations. Each cell reports the deviation, in percentage points, of a hardware platform's mean score from the mean across the three platforms for the corresponding task; positive and negative values indicate performance above and below the task mean, respectively. Srv-SMU, Srv-STF, and WS-SMU denote Server-SMU, Server-STF, and Workstation-SMU. Aggregates include only the agent configurations, models, and prompting strategies available on all three platforms.}
    \label{fig:rq1_hardware_heatmap}
\end{figure}

Although hardware is not expected to change task accuracy in the same direct way as model, prompt, or agentic configuration, platform-specific runtime differences and potential non-deterministic factors can still affect observed results. To avoid confounding hardware with model availability, Table~\ref{tab:rq1_hardware} and Figure~\ref{fig:rq1_hardware_heatmap} restrict the comparison to configurations that are available on all three hardware platforms, meaning that we exclude the GPT-OSS model. Under this balanced comparison, hardware differences are comparatively small. The largest task-level spreads are 2.5 percentage points for code generation and 2.0 percentage points for vulnerability detection, while technical debt detection, log parsing, and log analysis differ by less than one percentage point across hardware. Thus, hardware is not a major driver of RQ1 effectiveness compared with prompt strategy, model choice, and agentic configuration.

\begin{table*}[t]
\centering
\caption{RQ1 task-specific effectiveness across hardware platforms on the hardware-balanced subset. Values report mean $\pm$ standard deviation over the agent configurations, models, and prompting strategies available on all three platforms; GPT-OSS is excluded.}
\label{tab:rq1_hardware}
\scriptsize
\begin{tabular}{lr@{ $\pm$ }lr@{ $\pm$ }lr@{ $\pm$ }l}
\toprule
Task & \multicolumn{2}{c}{Server-SMU} & \multicolumn{2}{c}{Server-STF} & \multicolumn{2}{c}{Workstation-SMU} \\
\cmidrule(lr){2-3} \cmidrule(lr){4-5} \cmidrule(lr){6-7}
 & mean & std & mean & std & mean & std \\
\midrule
Code Generation (Pass@1) & \textbf{96.4} & \textbf{4.7} & 94.1 & 15.9 & 93.9 & 10.9 \\
Technical Debt Detection (Macro F1) & \textbf{38.4} & \textbf{11.7} & 38.0 & 11.6 & 38.1 & 11.5 \\
Vulnerability Detection (F1-score) & \textbf{46.5} & \textbf{17.1} & 44.5 & 18.8 & 45.9 & 17.2 \\
Log Parsing (Parsing Accuracy) & 27.9 & 31.2 & \textbf{28.6} & \textbf{31.9} & 28.2 & 31.6 \\
Log Analysis (F1-score) & \textbf{6.8} & \textbf{3.1} & 6.6 & 3.0 & 6.1 & 2.5 \\
\bottomrule
\end{tabular}
\end{table*}

\subsubsection{Effect of prompt strategy.}
Table~\ref{tab:rq1_prompt} shows that prompt strategy has a strongly task-dependent effect. Zero-shot prompting performs best for code generation, technical debt detection, and log analysis, while few-shot prompting performs marginally better for vulnerability detection and dramatically better for log parsing. The log parsing result is the task where prompt strategy has the largest effect: few-shot prompting increases parsing accuracy from 3.4\% to 54.3\%. This suggests that, for output-format-sensitive tasks, concrete examples are much more valuable than additional agentic orchestration. By contrast, code generation is better in the zero-shot setting (97.3\% versus 92.7\%), indicating that examples can sometimes constrain or distract the model rather than help.

\begin{table*}[t]
\centering
\caption{RQ1 effect of prompting strategy on task-specific effectiveness. Values report mean $\pm$ standard deviation for zero-shot and three-example few-shot prompting, averaged over agent configurations, models, and hardware platforms.}
\label{tab:rq1_prompt}
\scriptsize
\begin{tabular}{lr@{ $\pm$ }lr@{ $\pm$ }l}
\toprule
Task & \multicolumn{2}{c}{Zero-shot} & \multicolumn{2}{c}{Few-shot} \\
\cmidrule(lr){2-3} \cmidrule(lr){4-5}
 & mean & std & mean & std \\
\midrule
Code Generation (Pass@1) & \textbf{97.3} & \textbf{6.4} & 92.7 & 14.0 \\
Technical Debt Detection (Macro F1) & \textbf{40.2} & \textbf{12.2} & 38.7 & 12.4 \\
Vulnerability Detection (F1-score) & 45.8 & 18.0 & \textbf{46.8} & \textbf{16.7} \\
Log Parsing (Parsing Accuracy) & 3.4 & 6.9 & \textbf{54.3} & \textbf{24.1} \\
Log Analysis (F1-score) & \textbf{7.2} & \textbf{3.2} & 5.7 & 2.2 \\
\bottomrule
\end{tabular}
\end{table*}

\subsubsection{Effect of LLM model.}
Model choice also has a larger and more task-specific effect than agentic configuration. As shown in Table~\ref{tab:rq1_model}, no single model is best across all tasks. G3-4B obtains the highest Pass@1 for code generation (99.8\%), GPT-OSS obtains the highest Macro F1 for technical debt detection (58.6\%), Q3-4B performs best for vulnerability detection (57.8\%), and Q2.5C-7B performs best for both log parsing (42.2\%) and log analysis (8.5\%). These differences suggest that task--model matching is more important than simply increasing the number of agents. The results also show that the common assumption that larger models perform better than smaller ones is not always true: 4B models provide the strongest local-model accuracy for code generation and vulnerability detection, while Q2.5C-7B leads the two log tasks.

\begin{table*}[t]
\centering
\caption{RQ1 task-specific effectiveness by LLM model. Values report mean $\pm$ standard deviation across agent configurations, prompting strategies, and available hardware platforms. Performance is measured using Pass@1 for code generation, Macro F1 for technical debt detection, F1-score for vulnerability detection and log analysis, and exact-match parsing accuracy for log parsing. GPT-OSS results include only Server-SMU.}
\label{tab:rq1_model}
\scriptsize
\begin{tabular}{lr@{ $\pm$ }lr@{ $\pm$ }lr@{ $\pm$ }lr@{ $\pm$ }lr@{ $\pm$ }lr@{ $\pm$ }l}
\toprule
Task & \multicolumn{2}{c}{CG-7B} & \multicolumn{2}{c}{DS-6.7B} & \multicolumn{2}{c}{G3-4B} & \multicolumn{2}{c}{GPT-OSS} & \multicolumn{2}{c}{Q2.5C-7B} & \multicolumn{2}{c}{Q3-4B} \\
\cmidrule(lr){2-3} \cmidrule(lr){4-5} \cmidrule(lr){6-7} \cmidrule(lr){8-9} \cmidrule(lr){10-11} \cmidrule(lr){12-13}
 & mean & std & mean & std & mean & std & mean & std & mean & std & mean & std \\
\midrule
Code Generation (Pass@1) & 95.6 & 4.7 & 88.8 & 19.8 & \textbf{99.8} & \textbf{0.5} & 97.9 & 1.9 & 98.0 & 3.0 & 91.8 & 12.8 \\
Technical Debt Detection (Macro F1) & 29.7 & 2.9 & 23.3 & 3.2 & 52.6 & 2.6 & \textbf{58.6} & \textbf{3.7} & 47.2 & 3.8 & 37.9 & 6.2 \\
Vulnerability Detection (F1-score) & 35.1 & 19.9 & 40.4 & 18.2 & 51.1 & 9.2 & 56.5 & 5.4 & 43.9 & 19.1 & \textbf{57.8} & \textbf{9.8} \\
Log Parsing (Parsing Accuracy) & 24.9 & 30.8 & 15.1 & 28.1 & 28.0 & 29.1 & 38.1 & 27.5 & \textbf{42.2} & \textbf{37.3} & 30.9 & 26.4 \\
Log Analysis (F1-score) & 7.0 & 3.7 & 5.2 & 1.3 & 5.6 & 0.6 & 5.1 & 1.5 & \textbf{8.5} & \textbf{3.4} & 6.4 & 3.0 \\
\bottomrule
\end{tabular}
\end{table*}

\begin{figure}
    \centering
    \includegraphics[width=1.0\linewidth]{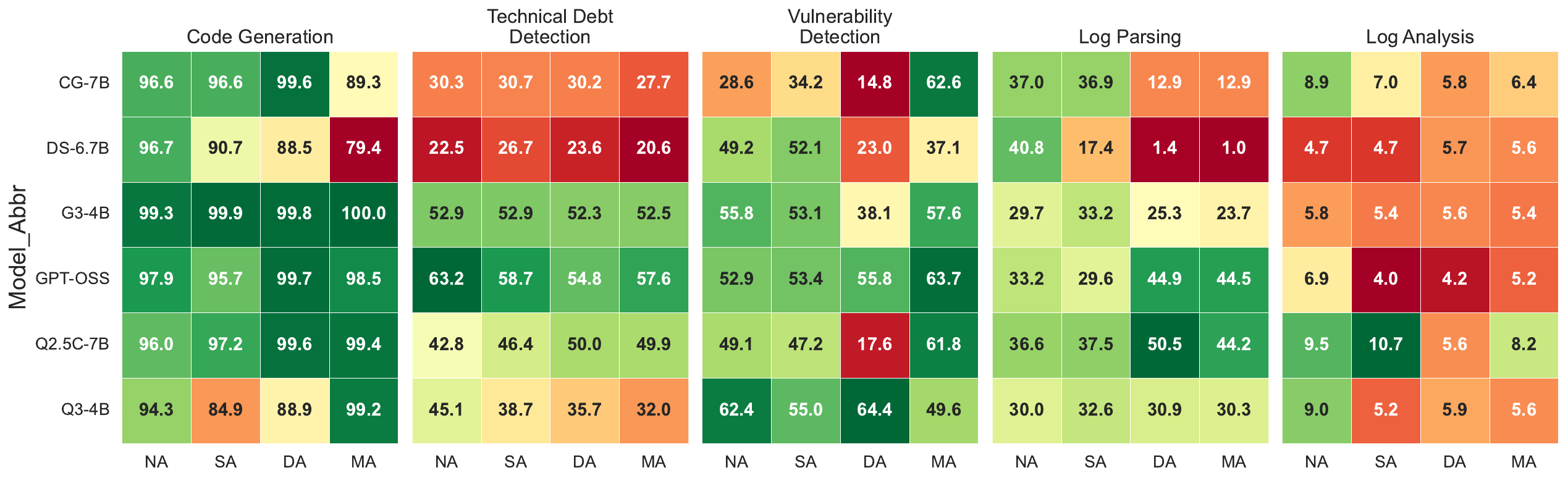}
     \vspace*{-0.90em}
    \caption{RQ1 task-specific effectiveness by LLM model and system configuration. Rows denote CodeGemma-7B (CG-7B), DeepSeek-Coder-6.7B (DS-6.7B), Gemma-3-4B (G3-4B), GPT-OSS-20B (GPT-OSS), Qwen2.5-Coder-7B (Q2.5C-7B), and Qwen3-4B (Q3-4B). Within each task panel, columns denote the non-agentic direct-query baseline (NA), single-agent (SA), dual-agent (DA), and four-agent multi-agent (MA) configurations. Each cell reports the corresponding task-specific performance metric as a percentage. Values are averaged over prompting strategies and the hardware platforms available for the corresponding model; GPT-OSS includes only Server-SMU measurements. Higher values indicate better performance.}
    \label{fig:rq1_model}
\end{figure}

Figure~\ref{fig:rq1_model} complements the model-level table (Table~\ref{tab:rq1_model}) by showing that the effect of agentic configuration is strongly model- and task-dependent. In vulnerability detection, MA performs strongly for several models, while DA remains highly model-dependent: it performs well with Q3-4B but is much weaker for several other models.

\subsubsection*{Task-specific deep dives}
The remainder of this subsection looks more closely at two tasks whose accuracy metrics allow a finer-grained view than the others: technical debt detection (per smell class) and log parsing (per similarity metric).

\subsubsection{Per-class analysis: Technical Debt Detection.}
For Technical Debt Detection, Macro F1 is used as the main comparison metric because it gives equal weight to each smell class. However, the per-class results reveal important differences that are hidden by the aggregate score. Table~\ref{tab:rq1_td_per_class_overall} shows that Long Method and No Smell are detected most reliably, with mean F1-scores of 62.7\% and 55.8\%, respectively. Data Class is more difficult (40.6\%), Blob is lower still (30.0\%), and Feature Envy is the main failure mode, with a mean F1-score of only 8.1\%.

\begin{table*}[t]
\centering
\caption{RQ1 technical-debt-specific per-class results: overall F1 score for each smell class, together with the best agent configuration, model, and prompting strategy. NA denotes the non-agentic direct-query baseline, SA the single-agent configuration, DA the dual-agent configuration, and MA the four-agent multi-agent configuration. Values in parentheses report the corresponding per-class F1 score (\%).}
\label{tab:rq1_td_per_class_overall}
\scriptsize
\begin{tabular}{lrrlll}
\toprule
Smell Class & Mean F1 & Std & Best Agent & Best Model & Best Prompt \\
\midrule
No Smell & 55.8 & 14.7 & NA (57.4) & G3-4B (69.6) & Zero-shot (56.9) \\
Blob & 30.0 & 20.8 & NA (32.8) & GPT-OSS (61.2) & Zero-shot (32.2) \\
Data Class & 40.6 & 26.5 & MA (41.7) & GPT-OSS (75.3) & Few-shot (41.3) \\
Feature Envy & 8.1 & 7.2 & DA (10.1) & CG-7B (13.1) & Zero-shot (10.1) \\
Long Method & 62.7 & 22.1 & SA (66.1) & GPT-OSS (88.7) & Few-shot (63.8) \\
\bottomrule
\end{tabular}
\end{table*}

The TD-specific breakdowns by agent, model, and prompt strategy are reported in Appendix~\ref{app:td_per_class}. 
The best agentic configuration varies by smell class: NA is best for No Smell and Blob, MA is best for Data Class, DA is best for Feature Envy, and SA is best for Long Method. However, these differences are small relative to the standard deviations, so they do not support a general claim that more agentic TD detection is better. Model choice is more consequential: GPT-OSS is best for Blob, Data Class, and Long Method, while G3-4B is best for No Smell and CG-7B is best for Feature Envy. Prompting also varies by class, with zero-shot performing best for No Smell, Blob, and Feature Envy, and few-shot performing best for Data Class and Long Method. Overall, the TD task illustrates why aggregate scores should be interpreted carefully: a configuration can look competitive in Macro F1 while still failing almost completely on a specific smell type.

\subsubsection{Metric sensitivity: Log Parsing.}
The main RQ1 analysis uses parsing accuracy as the primary log parsing metric because it is the strictest measure: a predicted template only counts as correct if it exactly matches the ground truth. However, this exact-match view can hide partial template quality. Table~\ref{tab:rq1_log_parsing_metrics} therefore reports two additional similarity metrics for log parsing: edit similarity and LCS similarity.

\begin{table*}[t]
\centering
\caption{RQ1 log parsing metric sensitivity: exact-match parsing accuracy and partial template similarity metrics (mean $\pm$ std in \%). The first four columns compare agent configurations using the main RQ1 agent aggregation; the final two columns compare prompt strategies using the main RQ1 prompt aggregation.}
\label{tab:rq1_log_parsing_metrics}
\scriptsize
\begin{tabular}{lr@{ $\pm$ }lr@{ $\pm$ }lr@{ $\pm$ }lr@{ $\pm$ }lr@{ $\pm$ }lr@{ $\pm$ }l}
\toprule
Metric & \multicolumn{2}{c}{NA} & \multicolumn{2}{c}{SA} & \multicolumn{2}{c}{DA} & \multicolumn{2}{c}{MA} & \multicolumn{2}{c}{Zero-shot} & \multicolumn{2}{c}{Few-shot} \\
\cmidrule(lr){2-3} \cmidrule(lr){4-5} \cmidrule(lr){6-7} \cmidrule(lr){8-9} \cmidrule(lr){10-11} \cmidrule(lr){12-13}
 & mean & std & mean & std & mean & std & mean & std & mean & std & mean & std \\
\midrule
Parsing Accuracy & \textbf{34.5} & \textbf{36.6} & 31.2 & 34.1 & 27.6 & 28.6 & 26.1 & 26.6 & 3.4 & 6.9 & \textbf{54.3} & \textbf{24.1} \\
Edit Similarity & \textbf{69.6} & \textbf{25.4} & 64.0 & 27.1 & 63.3 & 28.7 & 65.0 & 25.5 & 46.7 & 16.1 & \textbf{80.8} & \textbf{22.4} \\
LCS Similarity & 90.4 & 9.3 & 89.2 & 10.0 & 90.5 & 7.3 & \textbf{90.6} & \textbf{7.9} & 84.8 & 9.7 & \textbf{94.7} & \textbf{3.8} \\
\bottomrule
\end{tabular}
\end{table*}

The additional metrics change the absolute interpretation of log parsing performance, but not the main configuration-level conclusion. Under exact-match parsing accuracy, NA remains the strongest agentic configuration (34.5\%), followed by SA (31.2\%), DA (27.6\%), and MA (26.1\%). With edit similarity, the scores are much higher, ranging from 63.3\% to 69.6\%, and LCS similarity is higher still, around 89--91\% for all agentic configurations. This indicates that many incorrect templates are not arbitrary failures: they often preserve substantial token-level or subsequence structure even when they fail exact matching. At the same time, the partial-credit metrics do not reveal a hidden benefit from more complex agentic orchestration.

The strongest pattern remains the prompt effect. Few-shot prompting improves all three log parsing metrics, increasing parsing accuracy from 3.4\% to 54.3\%, edit similarity from 46.7\% to 80.8\%, and LCS similarity from 84.8\% to 94.7\%. Thus, the broader conclusion is robust across metrics: for log parsing, concrete output examples matter more than adding agents. Parsing accuracy is therefore retained as the conservative main metric, while edit similarity and LCS similarity help explain how close the failed templates are to the expected structure.

\begin{tcolorbox}[boxsep=2pt,left=2pt,right=2pt,top=2pt,bottom=2pt]
\paragraph{\textbf{RQ1 Conclusion.}} More complex agentic configurations do not provide a general accuracy benefit across the evaluated software engineering tasks, but MA is a task-specific exception for vulnerability detection. NA or SA is sufficient, and often best, for most tasks, while DA remains weak for vulnerability detection and the log tasks. The strongest effectiveness drivers are task-specific: prompt examples are crucial for log parsing, model choice varies sharply by task, and technical debt detection requires class-level analysis because \textit{Feature Envy} remains difficult even when aggregate Macro F1 appears moderate. Hardware has a negligible effect on accuracy, and the hardware-balanced statistical analysis supports this descriptive pattern: agentic configuration has a significant accuracy effect only for vulnerability detection.
\end{tcolorbox}

\subsection{Energy Consumption (RQ2)}
\label{subsec:RQ2}

RQ2 investigates how the system configurations differ in energy consumption across tasks and hardware platforms. 
The tables report mean $\pm$ standard deviation across the three repetitions of each experiment (except for the GPT-OSS model, where each experiment was only run once).

\subsubsection{Effect of agentic configuration.}
\begin{figure}
    \centering
    \includegraphics[width=1.0\linewidth]{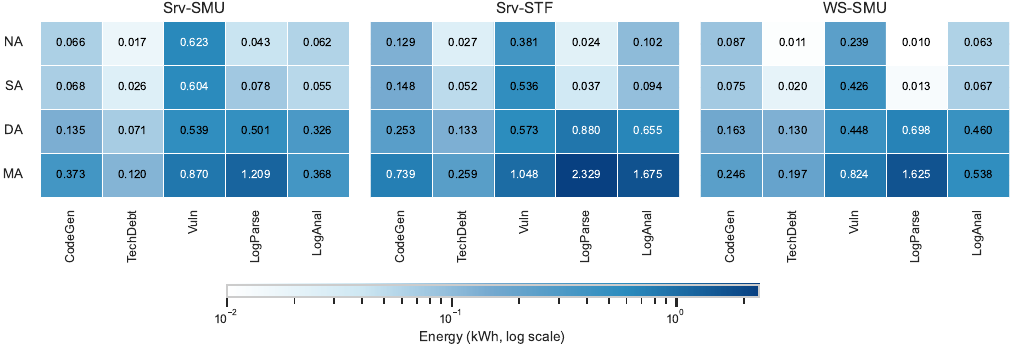}
    \vspace*{-0.90em}
    \caption{RQ2 mean energy consumption across software engineering tasks, agentic configurations, and hardware platforms. Rows represent the non-agentic direct-query baseline (NA), single-agent (SA), dual-agent (DA), and four-agent multi-agent (MA) configurations; columns represent code generation (CodeGen), technical debt detection (TechDebt), vulnerability detection (Vuln), log parsing (LogParse), and log analysis (LogAnal). Srv-SMU, Srv-STF, and WS-SMU denote Server-SMU, Server-STF, and Workstation-SMU, respectively. Each cell reports mean total pipeline energy consumption in kWh, aggregated over the available models, prompting strategies, and repetitions for that task--configuration--hardware combination. Color intensity is mapped logarithmically to accommodate the large range of energy values, whereas the cell annotations report the untransformed means.}
    \label{fig:rq2_energy}
\end{figure}

Figure~\ref{fig:rq2_energy} compares energy consumption across tasks, hardware platforms, and agentic configurations, while Table~\ref{tab:rq2_energy_average} summarizes the corresponding hardware-level and overall averages. The dominant pattern is that energy consumption generally increases with agentic complexity. Averaged over all tasks, models, prompts, and hardware platforms, as shown in the Overall row in Table~\ref{tab:rq2_energy_average}, SA consumes 1.20$\times$ as much energy as NA, DA consumes 3.06$\times$ as much, and MA consumes 6.36$\times$ as much. This increase is especially visible for the log tasks: for log parsing, MA reaches 1.209~kWh on Server-SMU, 2.329~kWh on Server-STF, and 1.625~kWh on Workstation-SMU, whereas NA remains below 0.05~kWh on all three platforms.

\begin{table*}[t]
\centering
\caption{RQ2: Average energy consumption (kWh) by hardware platform and agent configuration (mean $\pm$ std). Hardware rows average the task-level means shown in Figure~\ref{fig:rq2_energy}; the Overall row averages over all task, model, prompt, and hardware configurations. Lowest agent mean per row is bold.}
\label{tab:rq2_energy_average}
\scriptsize
\begin{tabular}{lr@{ $\pm$ }lr@{ $\pm$ }lr@{ $\pm$ }lr@{ $\pm$ }lr@{ $\pm$ }l}
\toprule
Hardware & \multicolumn{2}{c}{NA} & \multicolumn{2}{c}{SA} & \multicolumn{2}{c}{DA} & \multicolumn{2}{c}{MA} & \multicolumn{2}{c}{Avg.} \\
\cmidrule(lr){2-3} \cmidrule(lr){4-5} \cmidrule(lr){6-7} \cmidrule(lr){8-9} \cmidrule(lr){10-11}
 & mean & std & mean & std & mean & std & mean & std & mean & std \\
\midrule
Server-SMU & \textbf{0.162} & \textbf{0.006} & 0.166 & 0.012 & 0.315 & 0.031 & 0.588 & 0.090 & 0.308 & 0.048 \\
Server-STF & \textbf{0.132} & \textbf{0.010} & 0.173 & 0.007 & 0.499 & 0.021 & 1.210 & 1.714 & 0.504 & 0.857 \\
Workstation-SMU & \textbf{0.082} & \textbf{0.014} & 0.120 & 0.022 & 0.380 & 0.020 & 0.686 & 0.114 & 0.317 & 0.059 \\
Overall & \textbf{0.128} & \textbf{0.011} & 0.154 & 0.015 & 0.392 & 0.025 & 0.813 & 0.993 & 0.372 & 0.497 \\
\bottomrule
\end{tabular}
\end{table*}

The lowest-energy configuration is usually lightweight. Across the 15 task--hardware pairs in Figure~\ref{fig:rq2_energy}, NA is the lowest-energy option in 11 cases, SA in three cases, and DA in one case; MA is never the lowest-energy configuration. The aggregate values in Table~\ref{tab:rq2_energy_average} confirm that the overall energy burden is concentrated in the heavier DA and MA configurations rather than being evenly distributed across all setups.

In light of the results of RQ1, this is important because heavier agentic workflows rarely provide a corresponding general performance gain. Vulnerability detection is the main exception: MA gives the highest average F1-score for that task, but it also remains substantially more energy-intensive than NA and SA. DA, by contrast, remains less accurate than NA and SA while still increasing energy on most hardware platforms.

While the main pattern shows that adding multiple agents typically increases energy substantially, there are exceptions: SA is slightly lower than NA for code generation on Workstation-SMU and for log analysis on both servers, while DA is lowest for vulnerability detection on Server-SMU. These cases may arise when the agent orchestration produces shorter or more focused outputs than the unguided baseline, so that the reduced generation cost offsets the overhead of agent coordination.

\subsubsection{Effect of hardware platform.}
Hardware affects energy consumption, but the effect depends strongly on the agentic configuration, as summarized in Table~\ref{tab:rq2_energy_average}. For lightweight configurations, the differences between platforms are relatively small: NA ranges from 0.082~kWh on Workstation-SMU to 0.162~kWh on Server-SMU, and SA ranges from 0.120~kWh on Workstation-SMU to 0.173~kWh on Server-STF. For heavier configurations, the hardware gap widens. MA averages 0.588~kWh on Server-SMU, 0.686~kWh on Workstation-SMU, and 1.210~kWh on Server-STF. This suggests that hardware should not be treated as a simple constant multiplier. Hardware comparisons are most important when evaluating more complex agentic workflows; for lightweight configurations, architectural and model choices often dominate.

\begin{figure}
    \centering
    \includegraphics[width=0.73\linewidth]{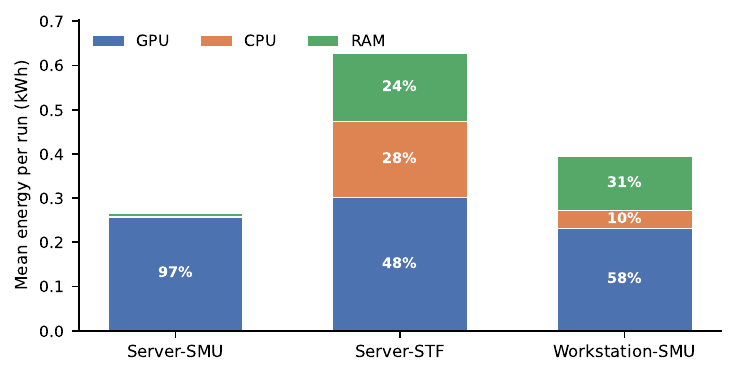}
    \vspace*{-0.90em}
    \caption{RQ2 mean process-attributed energy per run, decomposed into GPU, CPU, and RAM components for the three hardware platforms using the hardware-balanced subset of configurations (GPT-OSS excluded). Server-SMU is equipped with eight NVIDIA H100 GPUs with 80~GB HBM3 each, Server-STF with two NVIDIA RTX~6000 Ada GPUs with 48~GB each, and Workstation-SMU with one NVIDIA RTX~A2000 GPU with 12~GB. Bar height reports mean total pipeline energy in kWh, while segment labels report each component's share of the measured total. These percentages represent CodeCarbon energy attribution and should not be interpreted as device-utilization rates.}
    \label{fig:rq2_component_breakdown}
\end{figure}

The composition of that energy also differs markedly across platforms, as shown in Figure~\ref{fig:rq2_component_breakdown}. On Server-SMU, energy is almost entirely GPU-bound (97\%), with negligible CPU and RAM contributions. On Server-STF and Workstation-SMU, by contrast, CPU and RAM together account for roughly half of the measured energy (52\% and 42\%, respectively). The relative utilization of compute components is therefore hardware-specific rather than fixed, which reinforces that hardware cannot be modeled as a single constant multiplier on a GPU-only cost.

\subsubsection{Effect of prompt strategy.}
Table~\ref{tab:rq2_prompt_absolute} shows that prompt strategy has a task-specific energy effect. Few-shot prompting is not uniformly more expensive, despite the longer input prompt. For technical debt detection, few-shot uses less energy than zero-shot on all three hardware platforms, with Few/Zero energy ratios between 0.63 and 0.66. Log parsing shows the same pattern: few-shot consumes 24--41\% less energy than zero-shot across hardware platforms. This is notable because few-shot prompting also substantially improves log parsing accuracy in RQ1, making it one of the clearest cases where better performance and lower energy align.

\begin{table*}[t]
\centering
\caption{RQ2 effect of prompting strategy on total pipeline energy consumption by task and hardware platform. Values report mean $\pm$ standard deviation in kWh per run for zero-shot and three-example few-shot prompting, aggregated over the available models and agent configurations.}
\label{tab:rq2_prompt_absolute}
\scriptsize
\begin{tabular}{llr@{ $\pm$ }lr@{ $\pm$ }l}
\toprule
Task & Hardware & \multicolumn{2}{c}{Zero-shot} & \multicolumn{2}{c}{Few-shot} \\
\cmidrule(lr){3-4} \cmidrule(lr){5-6}
 &  & mean & std & mean & std \\
\midrule
\multirow{3}{*}{Code Generation} & Server-SMU & \textbf{0.157} & \textbf{0.001} & 0.164 & 0.002 \\
 & Server-STF & \textbf{0.149} & \textbf{0.003} & 0.485 & 1.691 \\
 & Workstation-SMU & \textbf{0.089} & \textbf{0.044} & 0.196 & 0.047 \\
\hline
\multirow{3}{*}{Technical Debt Detection} & Server-SMU & 0.071 & 0.007 & \textbf{0.046} & \textbf{0.007} \\
 & Server-STF & 0.141 & 0.008 & \textbf{0.094} & \textbf{0.011} \\
 & Workstation-SMU & 0.110 & 0.027 & \textbf{0.069} & \textbf{0.013} \\
\hline
\multirow{3}{*}{Vulnerability Detection} & Server-SMU & 0.681 & 0.064 & \textbf{0.637} & \textbf{0.060} \\
 & Server-STF & \textbf{0.611} & \textbf{0.042} & 0.658 & 0.029 \\
 & Workstation-SMU & \textbf{0.464} & \textbf{0.040} & 0.504 & 0.019 \\
\hline
\multirow{3}{*}{Log Parsing} & Server-SMU & 0.520 & 0.102 & \textbf{0.396} & \textbf{0.007} \\
 & Server-STF & 1.009 & 0.053 & \textbf{0.626} & \textbf{0.006} \\
 & Workstation-SMU & 0.738 & 0.130 & \textbf{0.435} & \textbf{0.103} \\
\hline
\multirow{3}{*}{Log Analysis} & Server-SMU & \textbf{0.188} & \textbf{0.049} & 0.218 & 0.050 \\
 & Server-STF & 0.825 & 2.118 & \textbf{0.438} & \textbf{0.039} \\
 & Workstation-SMU & \textbf{0.273} & \textbf{0.020} & 0.291 & 0.017 \\
\bottomrule
\end{tabular}
\end{table*}

The opposite pattern appears for code generation, where few-shot prompting increases energy on all platforms, most strongly on Server-STF and Workstation-SMU. Vulnerability detection is close to neutral, with Few/Zero ratios near 1.0. Log analysis is hardware-dependent: few-shot is lower on Server-STF but slightly higher on Server-SMU and Workstation-SMU. These mixed results indicate that prompt length alone does not determine energy consumption. The generated output length, model behavior, and agent workflow dynamics can dominate the cost of the initial prompt.

\subsubsection{Effect of LLM model.}
Model choice has a large effect on energy consumption, often larger than the hardware differences observed for a fixed agentic configuration. Table~\ref{tab:rq2_model_absolute} shows that the lowest-energy model varies by task and hardware, but two models dominate the low-energy cases: CG-7B is the lowest-energy model in seven of the 15 task--hardware pairs, and Q3-4B is the lowest in six. G3-4B and Q2.5C-7B are each lowest in one case. DS-6.7B and GPT-OSS are never the lowest-energy models in this comparison. 

\begin{table*}[t]
\centering
\caption{RQ2 total pipeline energy consumption by task, LLM model, and hardware platform. Values report mean $\pm$ standard deviation in kWh per run, averaged over agent configurations, prompting strategies, and available repetitions. Dashes indicate unsupported model--hardware combinations. GPT-OSS was evaluated only once per condition on Server-SMU; therefore, its reported zero standard deviation does not indicate zero variability.}
\label{tab:rq2_model_absolute}
\scriptsize
\begin{tabular}{llr@{ $\pm$ }lr@{ $\pm$ }lr@{ $\pm$ }lr@{ $\pm$ }lr@{ $\pm$ }lr@{ $\pm$ }l}
\toprule
Task & Hardware & \multicolumn{2}{c}{CG-7B} & \multicolumn{2}{c}{DS-6.7B} & \multicolumn{2}{c}{G3-4B} & \multicolumn{2}{c}{GPT-OSS} & \multicolumn{2}{c}{Q2.5C-7B} & \multicolumn{2}{c}{Q3-4B} \\
\cmidrule(lr){3-4} \cmidrule(lr){5-6} \cmidrule(lr){7-8} \cmidrule(lr){9-10} \cmidrule(lr){11-12} \cmidrule(lr){13-14}
 &  & mean & std & mean & std & mean & std & mean & std & mean & std & mean & std \\
\midrule
\multirow{3}{*}{CodeGen} & Server-SMU & \textbf{0.045} & \textbf{0.001} & 0.230 & 0.001 & 0.062 & 0.001 & 0.438 & 0.000 & 0.061 & 0.000 & 0.127 & 0.002 \\
 & Server-STF & \textbf{0.090} & \textbf{0.002} & 0.449 & 0.002 & 0.129 & 0.008 & --- & --- & 0.126 & 0.006 & 0.792 & 2.673 \\
 & Workstation-SMU & 0.071 & 0.000 & 0.420 & 0.002 & \textbf{0.018} & \textbf{0.005} & --- & --- & 0.100 & 0.000 & 0.104 & 0.102 \\
\hline
\multirow{3}{*}{TechDebt} & Server-SMU & \textbf{0.016} & \textbf{0.000} & 0.210 & 0.001 & 0.018 & 0.010 & 0.070 & 0.000 & 0.020 & 0.000 & 0.017 & 0.011 \\
 & Server-STF & 0.034 & 0.000 & 0.443 & 0.013 & 0.047 & 0.015 & --- & --- & 0.032 & 0.001 & \textbf{0.032} & \textbf{0.007} \\
 & Workstation-SMU & 0.027 & 0.000 & 0.340 & 0.045 & 0.032 & 0.014 & --- & --- & \textbf{0.021} & \textbf{0.000} & 0.027 & 0.001 \\
\hline
\multirow{3}{*}{Vuln} & Server-SMU & \textbf{0.135} & \textbf{0.010} & 0.480 & 0.002 & 0.200 & 0.137 & 2.487 & 0.000 & 0.271 & 0.001 & 0.382 & 0.021 \\
 & Server-STF & \textbf{0.297} & \textbf{0.002} & 1.114 & 0.017 & 0.596 & 0.075 & --- & --- & 0.498 & 0.003 & 0.667 & 0.025 \\
 & Workstation-SMU & \textbf{0.239} & \textbf{0.001} & 0.900 & 0.003 & 0.334 & 0.030 & --- & --- & 0.380 & 0.049 & 0.567 & 0.040 \\
\hline
\multirow{3}{*}{LogParse} & Server-SMU & 0.018 & 0.000 & 0.364 & 0.150 & 0.109 & 0.035 & 0.551 & 0.000 & 1.689 & 0.050 & \textbf{0.014} & \textbf{0.011} \\
 & Server-STF & 0.039 & 0.000 & 0.657 & 0.008 & 0.085 & 0.014 & --- & --- & 3.283 & 0.083 & \textbf{0.025} & \textbf{0.001} \\
 & Workstation-SMU & 0.029 & 0.000 & 0.485 & 0.003 & 0.027 & 0.002 & --- & --- & 2.374 & 0.263 & \textbf{0.017} & \textbf{0.000} \\
\hline
\multirow{3}{*}{LogAnal} & Server-SMU & 0.142 & 0.002 & 0.294 & 0.003 & 0.197 & 0.094 & 0.347 & 0.000 & 0.148 & 0.001 & \textbf{0.089} & \textbf{0.059} \\
 & Server-STF & \textbf{0.310} & \textbf{0.001} & 0.735 & 0.008 & 0.437 & 0.037 & --- & --- & 0.316 & 0.001 & 1.360 & 3.349 \\
 & Workstation-SMU & 0.242 & 0.002 & 0.486 & 0.002 & 0.219 & 0.025 & --- & --- & 0.253 & 0.002 & \textbf{0.209} & \textbf{0.032} \\
\bottomrule
\end{tabular}
\vspace{1mm}
\raggedright
\footnotesize\textit{Note.} GPT-OSS was evaluated with a single run on Server-SMU only; reported std = 0.000 reflects the absence of repeated trials rather than zero variability.
\end{table*}

\begin{figure}
    \centering
    \includegraphics[width=0.82\linewidth]{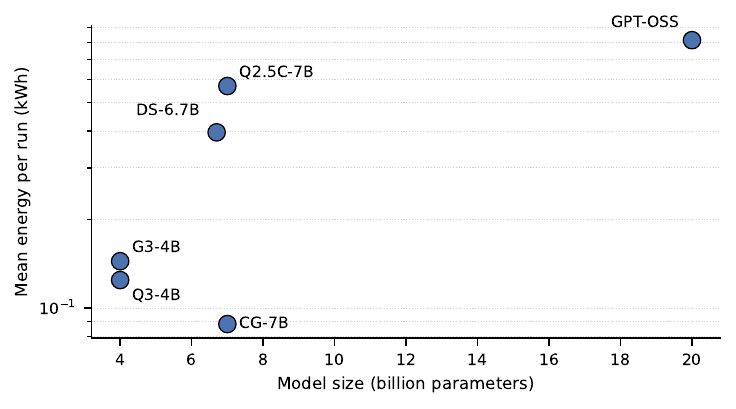}
    \vspace*{-0.90em}
    \caption{RQ2 relationship between model size and mean total pipeline energy consumption on Server-SMU, the only platform on which all six models were evaluated. Each point represents a model's mean energy per run, averaged across the five tasks, four agentic configurations, and two prompting strategies; the vertical axis is logarithmic, while the plotted values remain in kWh. Energy does not scale monotonically with parameter count: CG-7B is the lowest-energy model, whereas the similarly sized Q2.5C-7B consumes 6.4$\times$ more energy on average. GPT-OSS was evaluated once per condition, while the other models used three repetitions.}
    \label{fig:rq2_model_size_energy}
\end{figure}

The magnitude of the model effect can be large. For log parsing, Q3-4B is consistently the lowest-energy model, while Q2.5C-7B consumes more than two orders of magnitude more energy than the best model in several task--hardware pairs. For vulnerability detection, CG-7B is consistently the lowest-energy model across hardware platforms. However, the energy ranking should not be read as a performance recommendation by itself. In RQ1, CG-7B is not the best-performing model for any task-level model average, whereas Q2.5C-7B performs best for the two log tasks and GPT-OSS performs best for technical debt detection. These models can therefore be attractive from an effectiveness perspective even when they are not the lowest-energy choice. Conversely, Q3-4B is often energy-efficient and also performs best for vulnerability detection, making it a stronger candidate for that task than energy alone would suggest. The results also challenge the common assumption that a model with fewer parameters is generally the more energy-efficient option.

Figure~\ref{fig:rq2_model_size_energy} makes this explicit by plotting mean energy against model size on Server-SMU, the only platform on which all six models were evaluated. Energy does not grow monotonically with parameter count: the two 7B models differ by a factor of $6.4\times$ (CG-7B at 0.088~kWh versus Q2.5C-7B at 0.569~kWh), and CG-7B is in fact the lowest-energy model overall, below both 4B models. Only the 20B GPT-OSS model stands clearly apart at the high end, so larger size is associated with higher energy only at the extreme of the range, not within the cluster of small models.

These results reinforce that sustainable configuration design cannot be reduced to choosing fewer agents, a specific hardware platform, or the lowest-energy model in isolation. Model selection is a significant energy driver, but the relevant decision criterion is energy for sufficient task performance. The model-level table is therefore best interpreted as identifying candidate low-energy models that should be cross-checked against the RQ1 performance results and the RQ4 trade-off analysis.

\begin{tcolorbox}[boxsep=2pt,left=2pt,right=2pt,top=2pt,bottom=2pt]
\paragraph{\textbf{RQ2 Conclusion.}} Energy consumption is primarily driven by agentic complexity and model choice, with hardware effects becoming most pronounced for heavier DA and MA workflows. NA is the lowest-energy configuration in most task--hardware pairs, and MA is never the lowest-energy option. Prompt strategy has a task-specific effect: few-shot prompting reduces energy for technical debt detection and log parsing, but increases it for code generation. When read together with RQ1, the most defensible energy-efficient designs are not simply the lowest-energy configurations, but lightweight, task-aware configurations that retain sufficient performance.
\end{tcolorbox}

\subsection{Inference Latency Comparison (RQ3)}
\label{subsec:RQ3}

RQ3 examines the end-to-end inference latency of the evaluated configurations. Latency includes the complete time from issuing the prompt to receiving the final output, including model inference and any agent orchestration overhead. As in RQ2, the tables report mean $\pm$ standard deviation across repeated runs for each configuration.

\subsubsection{Effect of agentic configuration.}
Table~\ref{tab:rq3_latency} presents the latency for each task--hardware pair across the four agentic configurations, while Table~\ref{tab:rq3_latency_average} summarizes the corresponding hardware-level and overall averages. The overall pattern closely mirrors the energy results in RQ2: latency increases with agentic complexity, but the magnitude is task-dependent. Averaged over all tasks, models, prompts, and hardware platforms, as shown in the Overall row in Table~\ref{tab:rq3_latency_average}, SA is 1.23$\times$ slower than NA, DA is 3.32$\times$ slower, and MA is 6.07$\times$ slower. The most extreme case is log parsing on Workstation-SMU, where MA takes 50\,186~seconds (nearly 14~hours) compared with 313~seconds for NA---a 160$\times$ slowdown.

\begin{table*}[t]
\centering
\caption{RQ3 end-to-end latency by task, hardware platform, and LLM system configuration. Values report mean $\pm$ standard deviation in seconds, aggregated over available models, prompting strategies, and repetitions. NA, SA, DA, and MA denote the non-agentic, single-agent, dual-agent, and four-agent configurations, respectively; the lowest mean in each task--hardware row is shown in bold.}
\label{tab:rq3_latency}
\scriptsize
\begin{tabular}{llr@{ $\pm$ }lr@{ $\pm$ }lr@{ $\pm$ }lr@{ $\pm$ }l}
\toprule
Task & Hardware & \multicolumn{2}{c}{NA} & \multicolumn{2}{c}{SA} & \multicolumn{2}{c}{DA} & \multicolumn{2}{c}{MA} \\
\cmidrule(lr){3-4} \cmidrule(lr){5-6} \cmidrule(lr){7-8} \cmidrule(lr){9-10}
 &  & mean & std & mean & std & mean & std & mean & std \\
\midrule
\multirow{3}{*}{Code Generation} & Server-SMU & 726.7 & 15.8 & \textbf{712.0} & \textbf{9.8} & 1403.3 & 19.9 & 3935.5 & 36.5 \\
 & Server-STF & \textbf{955.6} & \textbf{6.5} & 1023.7 & 3.2 & 1655.8 & 55.3 & 2220.4 & 570.4 \\
 & Workstation-SMU & 2672.6 & 142.8 & \textbf{2307.0} & \textbf{1519.3} & 4978.8 & 561.3 & 7538.7 & 2300.5 \\
\hline
\multirow{3}{*}{Technical Debt Detection} & Server-SMU & \textbf{199.6} & \textbf{2.1} & 289.5 & 9.0 & 713.6 & 6.7 & 1175.9 & 11.1 \\
 & Server-STF & \textbf{222.2} & \textbf{2.4} & 367.9 & 1.8 & 951.4 & 94.5 & 1785.8 & 72.9 \\
 & Workstation-SMU & \textbf{345.8} & \textbf{10.3} & 629.4 & 140.6 & 3983.2 & 156.5 & 6086.4 & 1368.3 \\
\hline
\multirow{3}{*}{Vulnerability Detection} & Server-SMU & 6996.8 & 56.7 & 6592.2 & 35.1 & \textbf{5976.9} & \textbf{277.4} & 9790.7 & 331.4 \\
 & Server-STF & \textbf{2669.1} & \textbf{8.8} & 3564.7 & 19.2 & 3870.3 & 78.5 & 7450.1 & 676.0 \\
 & Workstation-SMU & \textbf{7330.6} & \textbf{220.9} & 13053.2 & 234.3 & 13734.4 & 875.9 & 24687.7 & 4695.7 \\
\hline
\multirow{3}{*}{Log Parsing} & Server-SMU & \textbf{526.7} & \textbf{2.6} & 942.7 & 243.2 & 4922.4 & 119.6 & 12475.5 & 3518.6 \\
 & Server-STF & \textbf{204.7} & \textbf{4.1} & 309.1 & 1.9 & 5467.7 & 15.1 & 14273.0 & 104.0 \\
 & Workstation-SMU & \textbf{313.1} & \textbf{3.5} & 391.1 & 3.1 & 21670.2 & 566.2 & 50185.8 & 7039.3 \\
\hline
\multirow{3}{*}{Log Analysis} & Server-SMU & 680.9 & 14.5 & \textbf{572.5} & \textbf{23.3} & 3755.4 & 363.8 & 4069.1 & 133.4 \\
 & Server-STF & 698.0 & 155.2 & \textbf{646.1} & \textbf{70.1} & 4497.4 & 82.3 & 5370.7 & 625.8 \\
 & Workstation-SMU & 2262.7 & 1560.3 & \textbf{2071.5} & \textbf{5.3} & 14104.7 & 267.7 & 16512.5 & 588.0 \\
\bottomrule
\end{tabular}
\end{table*}

\begin{table*}[t]
\centering
\caption{RQ3: Average inference latency (seconds) by hardware platform and agent configuration (mean $\pm$ std). Hardware rows average the task-level means shown in Figure~\ref{fig:rq3_latency}; the Overall row averages over all task, model, prompt, and hardware configurations. Lowest agent mean per row is bold.}
\label{tab:rq3_latency_average}
\scriptsize
\begin{tabular}{lr@{ $\pm$ }lr@{ $\pm$ }lr@{ $\pm$ }lr@{ $\pm$ }lr@{ $\pm$ }l}
\toprule
Hardware & \multicolumn{2}{c}{NA} & \multicolumn{2}{c}{SA} & \multicolumn{2}{c}{DA} & \multicolumn{2}{c}{MA} & \multicolumn{2}{c}{Avg.} \\
\cmidrule(lr){2-3} \cmidrule(lr){4-5} \cmidrule(lr){6-7} \cmidrule(lr){8-9} \cmidrule(lr){10-11}
 & mean & std & mean & std & mean & std & mean & std & mean & std \\
\midrule
Server-SMU & 1826.1 & 27.2 & \textbf{1821.8} & \textbf{110.5} & 3354.3 & 211.7 & 6289.3 & 1581.7 & 3322.9 & 799.9 \\
Server-STF & \textbf{949.9} & \textbf{69.6} & 1182.3 & 32.6 & 3288.5 & 70.9 & 6220.0 & 487.9 & 2910.2 & 249.5 \\
Workstation-SMU & \textbf{2585.0} & \textbf{707.6} & 3690.4 & 690.4 & 11694.3 & 547.5 & 21002.2 & 3977.7 & 9743.0 & 2067.6 \\
Overall & \textbf{1789.5} & \textbf{410.8} & 2205.9 & 404.1 & 5940.0 & 341.4 & 10865.5 & 2487.4 & 5200.2 & 1288.0 \\
\bottomrule
\end{tabular}
\end{table*}

As in the energy analysis, exceptions exist: For code generation and log analysis, SA is sometimes faster than NA (e.g., SA takes 712~s vs.\ NA's 727~s for code generation on Server-SMU, and 573~s vs.\ 681~s for log analysis on the same platform). These exceptions suggest that single-agent orchestration can occasionally streamline generation by producing more focused outputs.

\begin{figure}
    \centering
    \includegraphics[width=1.0\linewidth]{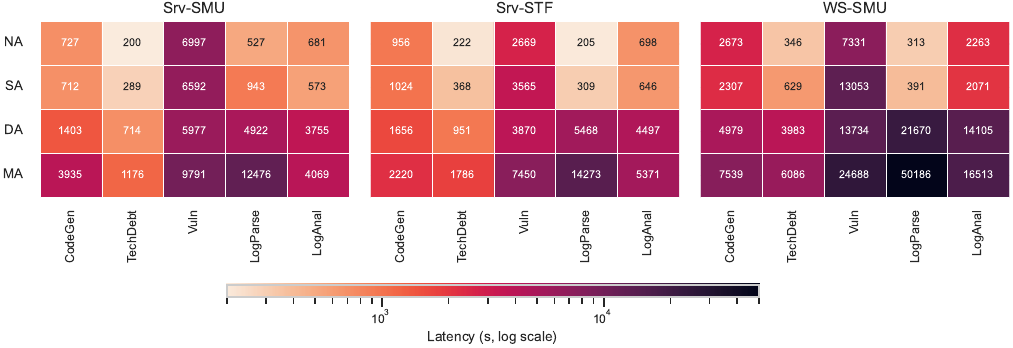}
    \vspace*{-1.20em}
    \caption{RQ3 mean end-to-end latency across software engineering tasks, agentic configurations, and hardware platforms. Rows represent the non-agentic direct-query baseline (NA), single-agent (SA), dual-agent (DA), and four-agent multi-agent (MA) configurations; columns represent code generation (CodeGen), technical debt detection (TechDebt), vulnerability detection (Vuln), log parsing (LogParse), and log analysis (LogAnal). Srv-SMU, Srv-STF, and WS-SMU denote Server-SMU, Server-STF, and Workstation-SMU, respectively. Each cell reports mean latency in seconds, averaged over the available models, prompting strategies, and repetitions for the corresponding task--configuration--hardware combination. Latency covers model loading, inference, and agent orchestration. Color intensity is mapped logarithmically to accommodate the wide range of values, whereas cell annotations report the untransformed arithmetic means.}
    \label{fig:rq3_latency}
\end{figure}

Figure~\ref{fig:rq3_latency} provides a compact task--hardware--agent view of the latency landscape, using the same heatmap format as the RQ2 energy analysis. The logarithmic color scale highlights the orders-of-magnitude range: the darkest cells (DA and MA for log parsing) are two orders of magnitude slower than the lightest cells (NA and SA for technical debt detection).

\subsubsection{Effect of hardware platform.}
Hardware has a large effect on latency, particularly for heavier configurations, as summarized in Table~\ref{tab:rq3_latency_average}. For NA and SA, Server-STF is generally the fastest platform (e.g., NA averages 950~s on Server-STF vs.\ 1\,826~s on Server-SMU and 2\,585~s on Workstation-SMU). For DA and MA, the hardware gap widens substantially: MA averages 6\,220~s on Server-STF but 21\,002~s on Workstation-SMU---a 3.4$\times$ difference driven by the workstation's slower GPU inference throughput for multi-round agent interactions. This suggests that hardware selection is especially important for complex agentic configurations, where latency differences between platforms grow from a factor of $\sim$2--3$\times$ for NA to $\sim$3.4$\times$ for MA.

\subsubsection{Effect of prompt strategy.}
Table~\ref{tab:rq3_prompt_absolute} shows that prompt strategy has a task-specific effect on latency, broadly consistent with the energy patterns in RQ2. Few-shot prompting reduces latency for technical debt detection (28--37\% faster across hardware) and log parsing (23--41\% faster), while increasing it for code generation on Server-STF and Workstation-SMU. Vulnerability detection is mixed, with few-shot lower on Server-SMU but higher on Server-STF and Workstation-SMU, whereas log analysis is consistently slower under few-shot prompting. These results confirm that, as with energy, prompt length is not the dominant factor; the model's generation behavior in response to examples matters more.

\begin{table*}[t]
\centering
\caption{RQ3 effect of prompting strategy on end-to-end latency by task and hardware platform. Values report mean $\pm$ standard deviation in seconds per run for zero-shot and three-example few-shot prompting, aggregated over the available models and agent configurations.}
\label{tab:rq3_prompt_absolute}
\scriptsize
\begin{tabular}{llr@{ $\pm$ }lr@{ $\pm$ }l}
\toprule
Task & Hardware & \multicolumn{2}{c}{Zero-shot} & \multicolumn{2}{c}{Few-shot} \\
\cmidrule(lr){3-4} \cmidrule(lr){5-6}
 &  & mean & std & mean & std \\
\midrule
\multirow{3}{*}{Code Generation} & Server-SMU & 1736.2 & 18.5 & \textbf{1652.6} & \textbf{26.4} \\
 & Server-STF & \textbf{1056.2} & \textbf{16.3} & 1871.5 & 404.9 \\
 & Workstation-SMU & \textbf{2730.7} & \textbf{1358.3} & 6017.9 & 1457.0 \\
\hline
\multirow{3}{*}{Technical Debt Detection} & Server-SMU & 707.9 & 9.1 & \textbf{481.4} & \textbf{6.7} \\
 & Server-STF & 968.2 & 3.8 & \textbf{695.5} & \textbf{84.3} \\
 & Workstation-SMU & 3383.2 & 884.4 & \textbf{2139.2} & \textbf{419.7} \\
\hline
\multirow{3}{*}{Vulnerability Detection} & Server-SMU & 7630.7 & 216.1 & \textbf{7047.6} & \textbf{221.2} \\
 & Server-STF & \textbf{4247.8} & \textbf{470.7} & 4529.3 & 100.9 \\
 & Workstation-SMU & \textbf{13948.3} & \textbf{3335.2} & 15454.7 & 580.2 \\
\hline
\multirow{3}{*}{Log Parsing} & Server-SMU & 5343.4 & 2494.2 & \textbf{4090.3} & \textbf{76.0} \\
 & Server-STF & 6270.8 & 73.4 & \textbf{3856.5} & \textbf{11.6} \\
 & Workstation-SMU & 22814.3 & 3804.2 & \textbf{13465.8} & \textbf{3234.8} \\
\hline
\multirow{3}{*}{Log Analysis} & Server-SMU & \textbf{2124.1} & \textbf{188.5} & 2414.9 & 199.8 \\
 & Server-STF & \textbf{2638.8} & \textbf{441.9} & 2967.3 & 135.7 \\
 & Workstation-SMU & \textbf{8468.7} & \textbf{877.5} & 9007.0 & 809.9 \\
\bottomrule
\end{tabular}
\end{table*}

\subsubsection{Effect of LLM model.}
Model choice has a large and consistent effect on latency. Table~\ref{tab:rq3_model_absolute} shows that CG-7B is the fastest model for code generation and vulnerability detection on both servers, while Q3-4B is fastest for log parsing on all platforms and for technical debt detection on the two servers. On Workstation-SMU, Q2.5C-7B is the fastest model for technical debt detection. DS-6.7B is often the slowest local model outside log parsing, reaching 23.6$\times$ the fastest latency for code generation on Workstation-SMU. The largest outlier is Q2.5C-7B on log parsing: it reaches 73\,544~s on Workstation-SMU ($\sim$20~hours), 136.5$\times$ slower than Q3-4B for the same task--hardware pair, likely due to extensive output generation for that task.

\begin{table*}[t]
\centering
\caption{RQ3 end-to-end latency by task, LLM model, and hardware platform. Values report mean $\pm$ standard deviation in seconds per run, averaged over agent configurations, prompting strategies, and available repetitions. Dashes indicate unsupported model--hardware combinations. GPT-OSS was evaluated once per condition on Server-SMU; therefore, its zero standard deviation does not imply zero variability.}
\label{tab:rq3_model_absolute}
\scriptsize
\begin{tabular}{llr@{ $\pm$ }lr@{ $\pm$ }lr@{ $\pm$ }lr@{ $\pm$ }lr@{ $\pm$ }lr@{ $\pm$ }l}
\toprule
Task & Hardware & \multicolumn{2}{c}{CG-7B} & \multicolumn{2}{c}{DS-6.7B} & \multicolumn{2}{c}{G3-4B} & \multicolumn{2}{c}{GPT-OSS} & \multicolumn{2}{c}{Q2.5C-7B} & \multicolumn{2}{c}{Q3-4B} \\
\cmidrule(lr){3-4} \cmidrule(lr){5-6} \cmidrule(lr){7-8} \cmidrule(lr){9-10} \cmidrule(lr){11-12} \cmidrule(lr){13-14}
 &  & mean & std & mean & std & mean & std & mean & std & mean & std & mean & std \\
\midrule
\multirow{3}{*}{CodeGen} & Server-SMU & \textbf{407.3} & \textbf{10.0} & 1952.5 & 20.4 & 878.4 & 43.1 & 4982.7 & 0.0 & 574.3 & 11.5 & 1371.0 & 9.5 \\
 & Server-STF & \textbf{627.3} & \textbf{5.7} & 2760.9 & 1.6 & 1085.9 & 68.5 & --- & --- & 859.9 & 15.0 & 1985.3 & 636.9 \\
 & Workstation-SMU & 2179.7 & 7.8 & 12862.7 & 66.7 & \textbf{545.6} & \textbf{159.3} & --- & --- & 3074.8 & 12.8 & 3208.5 & 3144.8 \\
\hline
\multirow{3}{*}{TechDebt} & Server-SMU & 183.9 & 3.2 & 1889.6 & 10.6 & 282.5 & 10.9 & 824.4 & 0.0 & 225.2 & 2.9 & \textbf{162.4} & \textbf{8.3} \\
 & Server-STF & 318.0 & 1.4 & 2765.0 & 5.4 & 495.7 & 121.5 & --- & --- & 298.1 & 2.4 & \textbf{282.4} & \textbf{54.8} \\
 & Workstation-SMU & 856.8 & 0.8 & 10446.6 & 1486.6 & 1002.0 & 430.6 & --- & --- & \textbf{656.7} & \textbf{2.2} & 844.0 & 18.8 \\
\hline
\multirow{3}{*}{Vuln} & Server-SMU & \textbf{1162.1} & \textbf{75.7} & 4051.7 & 18.5 & 3595.1 & 251.5 & 28162.0 & 0.0 & 2418.1 & 17.1 & 4646.0 & 411.6 \\
 & Server-STF & \textbf{2022.0} & \textbf{6.6} & 6888.3 & 7.4 & 4737.3 & 737.6 & --- & --- & 3243.5 & 11.5 & 5051.6 & 187.5 \\
 & Workstation-SMU & \textbf{7328.7} & \textbf{21.2} & 27545.9 & 95.1 & 10259.9 & 926.2 & --- & --- & 10948.8 & 5127.8 & 17424.0 & 1220.5 \\
\hline
\multirow{3}{*}{LogParse} & Server-SMU & 207.4 & 8.0 & 3745.3 & 3900.6 & 1394.1 & 306.1 & 7039.0 & 0.0 & 15743.4 & 508.1 & \textbf{171.8} & \textbf{25.8} \\
 & Server-STF & 352.9 & 2.5 & 4070.2 & 5.6 & 779.3 & 106.9 & --- & --- & 19880.0 & 48.4 & \textbf{235.8} & \textbf{2.0} \\
 & Workstation-SMU & 903.9 & 0.8 & 14850.8 & 95.9 & 862.3 & 59.1 & --- & --- & 73544.4 & 7894.8 & \textbf{538.7} & \textbf{3.7} \\
\hline
\multirow{3}{*}{LogAnal} & Server-SMU & \textbf{1214.2} & \textbf{34.3} & 2512.6 & 21.9 & 3267.4 & 430.6 & 3932.2 & 0.0 & 1317.5 & 13.6 & 1373.1 & 37.1 \\
 & Server-STF & 2019.6 & 4.5 & 4547.2 & 3.5 & 3537.4 & 274.9 & --- & --- & 1969.2 & 4.9 & \textbf{1942.0} & \textbf{677.2} \\
 & Workstation-SMU & 7406.1 & 62.6 & 14895.9 & 58.5 & \textbf{6739.4} & \textbf{763.6} & --- & --- & 7775.4 & 54.7 & 6872.4 & 1723.7 \\
\bottomrule
\end{tabular}
\vspace{1mm}
\raggedright
\footnotesize\textit{Note.} GPT-OSS was evaluated with a single run on Server-SMU only; reported std = 0.000 reflects the absence of repeated trials rather than zero variability.
\end{table*}

\begin{figure*}[t]
  \centering
  \includegraphics[width=\textwidth]{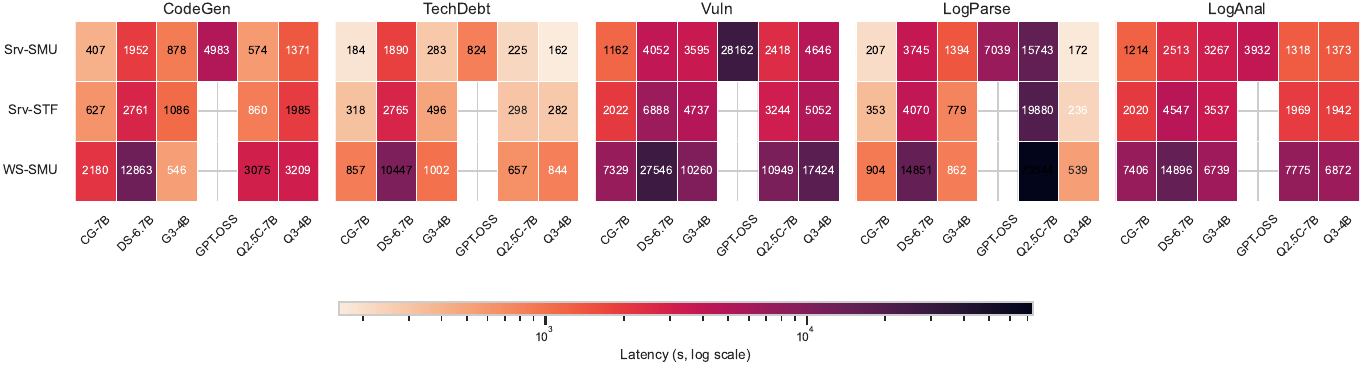}
  \vspace*{-1.20em}
  \caption{RQ3 mean end-to-end latency by LLM model and hardware platform, shown separately for code generation (CodeGen), technical debt detection (TechDebt), vulnerability detection (Vuln), log parsing (LogParse), and log analysis (LogAnal). Rows denote Server-SMU (Srv-SMU), Server-STF (Srv-STF), and Workstation-SMU (WS-SMU); columns denote CodeGemma-7B (CG-7B), DeepSeek-Coder-6.7B (DS-6.7B), Gemma-3-4B (G3-4B), GPT-OSS-20B (GPT-OSS), Qwen2.5-Coder-7B (Q2.5C-7B), and Qwen3-4B (Q3-4B). Each cell reports mean latency in seconds, averaged over agent configurations, prompting strategies, and available repetitions. Latency includes model loading, inference, and agent orchestration. Color intensity is mapped logarithmically, whereas cell annotations report the untransformed arithmetic means. Blank cells indicate unsupported model--hardware combinations; GPT-OSS was evaluated only on Server-SMU because of its memory requirements.}
  \label{fig:rq3_model_hardware_latency}
\end{figure*}

These model-level latency differences are generally larger than the hardware differences, reinforcing the RQ2 finding that model selection is a more impactful configuration lever than hardware choice. The fastest models (CG-7B, Q3-4B) tend to also be among the lowest-energy models in RQ2, because latency and energy are tightly coupled for local inference. Figure~\ref{fig:rq3_model_hardware_latency} visualizes the model--hardware interaction across tasks: DS-6.7B's slowdown on Workstation-SMU and Q2.5C-7B's pathological behavior on log parsing are clearly visible, as is the fact that GPT-OSS could only be evaluated on Server-SMU due to memory requirements (blank cells).

\subsubsection{Energy--latency relationship.}
\begin{figure}
    \centering
    \includegraphics[width=0.70\linewidth]{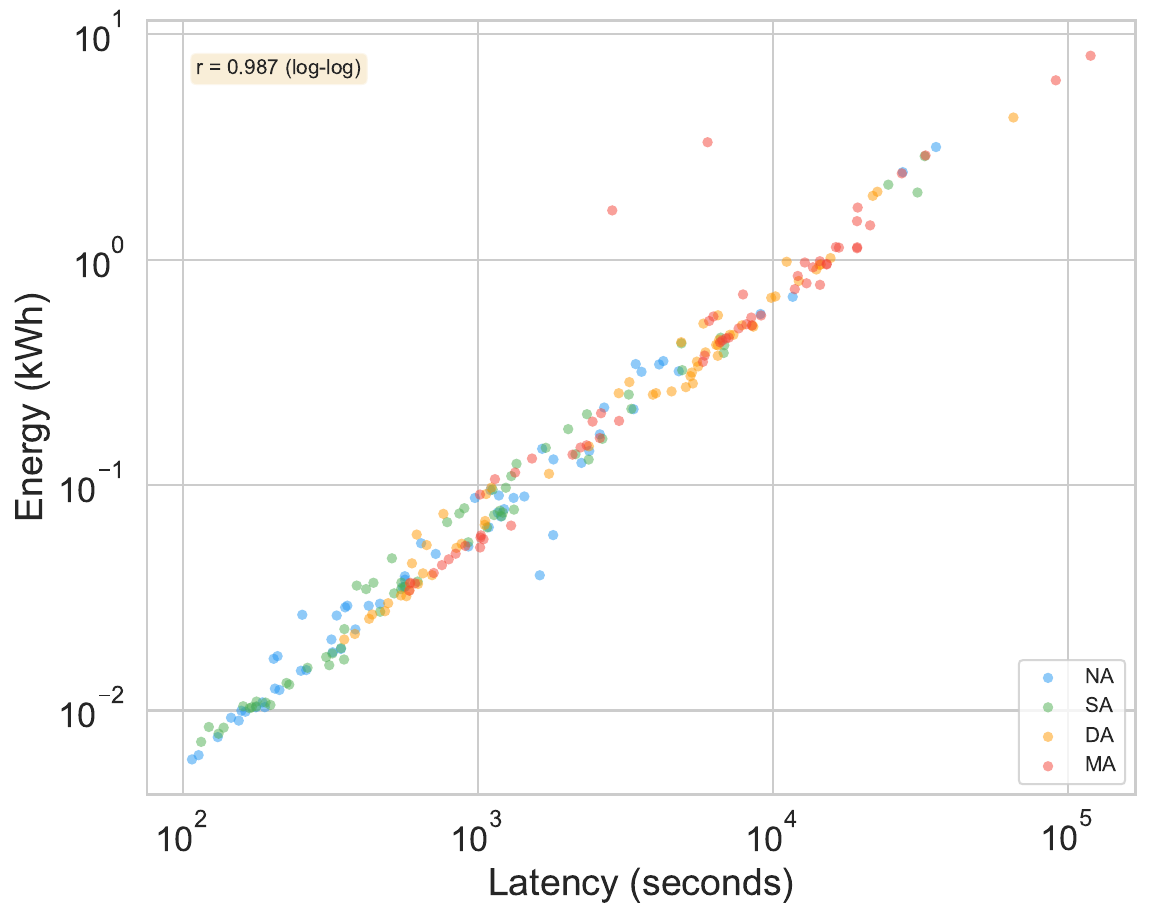}
    \vspace*{-1.20em}
    \caption{RQ3 relationship between total pipeline energy consumption and end-to-end latency across the hardware-balanced configurations. Each point represents one task--model--prompt--agent configuration, with energy measured in kWh and latency in seconds. Point color denotes the non-agentic baseline (NA), single-agent (SA), dual-agent (DA), or four-agent multi-agent (MA) configuration. Both axes use logarithmic scales, and the reported Pearson correlation is computed on the log-transformed values ($r=0.987$, $p<10^{-100}$). The strong positive association indicates that configurations with longer execution times generally also consume more energy; it does not by itself establish a causal relationship.}
    \label{fig:rq3_energy_latency}
\end{figure}

A distinctive feature of RQ3 is the near-perfect correlation between energy and latency. Across all hardware-balanced configurations, the Pearson correlation between log-transformed energy and log-transformed latency is $r = 0.987$ ($p < 10^{-100}$), as shown in Figure~\ref{fig:rq3_energy_latency}. This strong log--log relationship implies that configurations which are energy-expensive are also slow, and vice versa. In practical terms, this means that energy savings and latency savings are largely aligned: a configuration change that reduces one will almost always reduce the other. This tight coupling also means that the latency table can serve as a proxy for relative energy comparisons when energy measurements are unavailable.

\begin{tcolorbox}[boxsep=2pt,left=2pt,right=2pt,top=2pt,bottom=2pt]
\paragraph{\textbf{RQ3 Conclusion.}} Latency follows the same pattern as energy: it increases with agentic complexity, with MA being 6.07$\times$ slower than NA on average and up to 160$\times$ slower for individual task--hardware pairs. Hardware choice amplifies latency differences for heavier configurations, with Workstation-SMU being 3.4$\times$ slower than Server-STF for MA. Model choice has an even larger effect in some task--hardware pairs, with Q2.5C-7B being up to 136.5$\times$ slower than Q3-4B for log parsing. Energy and latency are strongly correlated ($r = 0.987$ in log--log space), meaning that optimizing for one effectively optimizes for the other. As with energy, prompt strategy is task-specific: few-shot reduces latency for technical debt detection and log parsing but can increase it for code generation and log analysis.
\end{tcolorbox}

\subsection{Trade-Off Analysis and Design Guidelines (RQ4)}
\label{subsec:RQ4}

RQ4 brings together the accuracy, energy, and latency dimensions examined in RQ1--RQ3 to identify configurations that achieve attractive multi-objective trade-offs. Since energy and latency are tightly correlated ($r = 0.987$, as shown in RQ3), the main visual analysis uses the two-dimensional accuracy--energy projection, while the configuration recommendations still account for latency when selecting a balanced Pareto-optimal setup.

\subsubsection{Pareto front analysis.}
\begin{figure}
    \centering
    \includegraphics[width=1.0\linewidth]{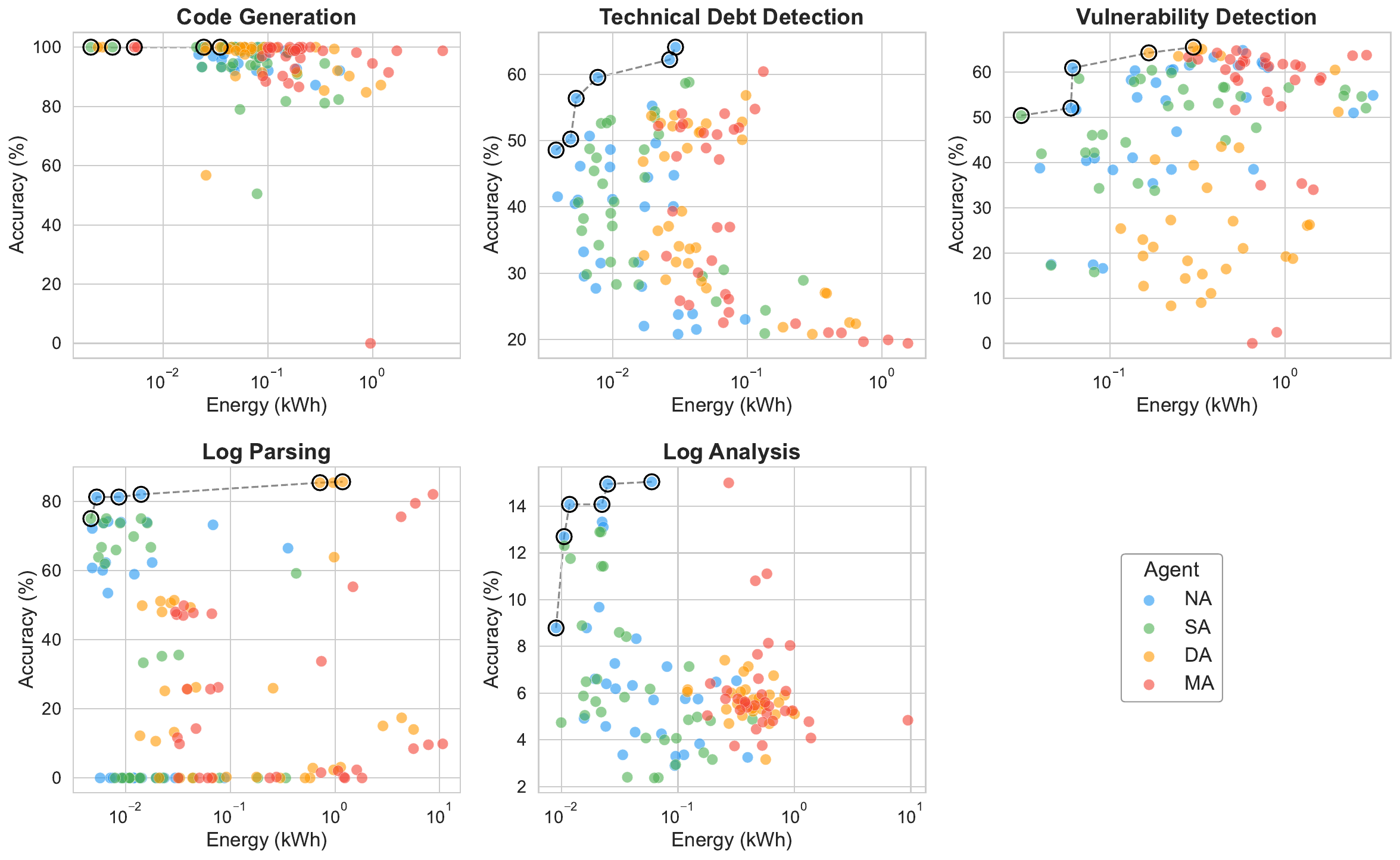}
    \vspace*{-1.10em}
    \caption{RQ4 task-level trade-off between task-specific performance and total pipeline energy consumption. Each point represents one model--agent--prompt configuration in the two-dimensional performance--energy projection; point color denotes the non-agentic direct-query baseline (NA), single-agent (SA), dual-agent (DA), or four-agent multi-agent (MA) configuration. Energy is reported in kWh on a logarithmic x-axis. The y-axis reports Pass@1 for code generation, Macro F1 for technical debt detection, F1-score for vulnerability detection and log analysis, and exact-match parsing accuracy for log parsing. Black circles identify Pareto-optimal configurations, for which no alternative achieves equal or higher performance with equal or lower energy, and dashed lines trace the resulting Pareto fronts. Latency is not represented in this projection but is included in the three-objective analysis reported in Table~\ref{tab:rq4_pareto_optimal}.}
    \label{fig:rq4_scatter}
\end{figure}

For each task--hardware pair, we compute the Pareto front over all configuration-level means (model $\times$ agent $\times$ prompt). In the two-dimensional projection, a configuration is Pareto-optimal if no other configuration achieves higher accuracy at equal or lower energy. Figure~\ref{fig:rq4_scatter} shows this accuracy--energy trade-off at the task level, with Pareto-optimal configurations highlighted. The corresponding Table~\ref{tab:rq4_pareto_optimal} uses the three-dimensional Pareto front (accuracy, energy, and latency) to select one balanced recommendation for each task--hardware pair.

For a configuration $c$ on the three-dimensional Pareto front of a given task--hardware pair, we compute a balanced score
\begin{equation}
\label{eq:balanced_score}
\mathrm{score}(c) = \tfrac{1}{3}\Big(\,\widetilde{a}(c) + \big(1 - \widetilde{e}(c)\big) + \big(1 - \widetilde{\ell}(c)\big)\Big),
\end{equation}
where $\widetilde{a}(c)$, $\widetilde{e}(c)$, and $\widetilde{\ell}(c)$ are the min--max-normalized accuracy, energy, and latency values of $c$ within the Pareto front of that task--hardware pair (each scaled to $[0,1]$, with accuracy higher-is-better and energy/latency lower-is-better). The configuration with the highest $\mathrm{score}(c)$ is reported in Table~\ref{tab:rq4_pareto_optimal}, with ties being broken by raw accuracy. This definition gives the three objectives equal weight on the Pareto front and avoids favoring the metric with the largest absolute range.

Across all 15 task--hardware pairs, the Pareto fronts contain a total of 66 configurations out of the full set of evaluated configurations for these pairs. Of the 66 Pareto-optimal configurations, 36 (55\%) use NA, 23 (35\%) use SA, 6 (9\%) use DA, and only 1 (2\%) uses MA. This confirms that lightweight configurations dominate the efficiency frontier. The few DA and MA Pareto points are task-specific: DA appears for log parsing and vulnerability detection, while MA appears once for log analysis. Thus, heavier agentic designs can occasionally survive the frontier when they improve accuracy, but they do not form a general efficient pattern.

Table~\ref{tab:rq4_pareto_optimal} reports the single best balanced Pareto configuration for each task--hardware pair. The column "Pareto front size" indicates how many Pareto-optimal configurations exist for that specific task--hardware pair. The selected configuration is chosen from the three-dimensional Pareto front by balancing normalized accuracy, energy, and latency, rather than by accuracy alone. For code generation, SA configurations achieve 100\% Pass@1 on all three hardware platforms while keeping energy and latency low, making them the clear balanced Pareto winners. Across the full table, NA is the balanced-best configuration in 11 of 15 task--hardware pairs; outside code generation, it is selected in 11 of 12 pairs.

\begin{table*}[t]
\centering
\caption{RQ4 balanced Pareto-optimal recommendation for each task--hardware pair, selected from the three-dimensional frontier using equally weighted normalized task performance, energy, and latency. Performance uses each task's primary effectiveness metric, and ``\# Pareto Configs'' reports the size of the corresponding frontier.}
\label{tab:rq4_pareto_optimal}
\scriptsize
\begin{tabular}{lllllrrrr}
\toprule
Task & Hardware & Agent & Model & Prompt & Accuracy & Energy (kWh) & Latency (s) & \# Pareto Configs \\
\midrule
\multirow{3}{*}{CodeGen} & Server-SMU & SA & DS-6.7B & Zero-shot & 100.0 & 0.0208 & 194.8 & 1 \\
 & Server-STF & SA & CG-7B & Zero-shot & 100.0 & 0.0448 & 333.2 & 4 \\
 & Workstation-SMU & SA & G3-4B & Few-shot & 100.0 & 0.0020 & 63.6 & 1 \\
\hline
\multirow{3}{*}{LogAnal} & Server-SMU & NA & Q3-4B & Few-shot & 14.9 & 0.0249 & 299.2 & 5 \\
 & Server-STF & NA & CG-7B & Zero-shot & 14.1 & 0.0222 & 201.2 & 6 \\
 & Workstation-SMU & NA & G3-4B & Zero-shot & 4.9 & 0.0156 & 494.8 & 6 \\
\hline
\multirow{3}{*}{LogParse} & Server-SMU & NA & DS-6.7B & Few-shot & 81.3 & 0.0053 & 62.1 & 3 \\
 & Server-STF & NA & DS-6.7B & Few-shot & 82.1 & 0.0140 & 110.6 & 4 \\
 & Workstation-SMU & NA & DS-6.7B & Few-shot & 81.3 & 0.0086 & 264.2 & 5 \\
\hline
\multirow{3}{*}{TechDebt} & Server-SMU & NA & Q3-4B & Zero-shot & 48.6 & 0.0038 & 69.4 & 7 \\
 & Server-STF & NA & Q3-4B & Zero-shot & 48.6 & 0.0096 & 96.5 & 2 \\
 & Workstation-SMU & NA & Q3-4B & Zero-shot & 50.2 & 0.0048 & 156.3 & 2 \\
\hline
\multirow{3}{*}{Vuln} & Server-SMU & NA & G3-4B & Zero-shot & 60.9 & 0.0613 & 1126.2 & 7 \\
 & Server-STF & SA & CG-7B & Zero-shot & 46.2 & 0.0908 & 699.8 & 7 \\
 & Workstation-SMU & NA & G3-4B & Few-shot & 51.7 & 0.0642 & 1982.1 & 6 \\
\bottomrule
\end{tabular}
\end{table*}

\begin{table*}[t]
\centering
\caption{RQ4 task-specific comparison of agentic configurations across effectiveness, energy consumption, and end-to-end latency. Each cell reports the best configuration, with its corresponding mean value in parentheses; ``Best Overall'' identifies the strongest combined trade-off across the three objectives.}
\label{tab:rq5_best_configs}
\scriptsize
\begin{tabular}{lcccc}
\toprule
Task & Best Accuracy & Lowest Energy & Lowest Latency & Best Overall \\
\midrule
Code Generation & NA (96.8) & NA (0.0924) & SA (1307.8) & NA \\
Technical Debt Detection & NA (42.8) & NA (0.0181) & NA (252.3) & NA \\
Vulnerability Detection & MA (55.4) & NA (0.4271) & NA (5748.7) & NA \\
Log Parsing & NA (34.5) & NA (0.0266) & NA (359.3) & NA \\
Log Analysis & NA (7.5) & SA (0.0712) & SA (1063.9) & SA \\
\bottomrule
\end{tabular}
\end{table*}

\subsubsection{Task-specific analysis.}
Table~\ref{tab:rq5_best_configs} complements the Pareto analysis by identifying the winning agentic configuration for each task across three objectives: highest accuracy, lowest energy, and lowest latency. The pattern is clear: NA is the best or tied-for-best agent in most cells, consistent with the Pareto front distribution reported. However, the exceptions reveal how task characteristics influence the balance between effectiveness and computational cost.
For vulnerability detection, MA achieves the highest average F1-score (55.4\%), whereas NA has the lowest energy consumption and latency. Although MA improves accuracy by 5.7 percentage points over NA, its substantially higher execution cost means that it appears on the balanced Pareto front for only one of the 15 task--hardware pairs. This task illustrates that measurable accuracy improvements from additional agents do not necessarily translate into the best overall cost--performance trade-off.
For code generation, SA is the balanced Pareto winner across all three hardware platforms, achieving 100\% Pass@1 while incurring comparable or occasionally lower latency than NA. Unlike the other tasks, the single-agent configuration preserves task accuracy while incurring little additional execution cost, making SA competitive with the NA configuration for this workload.
For technical debt detection, NA is the straightforward winner across all three objectives and all hardware platforms, with no tension between accuracy and cost. This task therefore provides the cleanest empirical support for the guideline to start non-agentic: additional orchestration neither improves accuracy nor reduces cost for this workload.
For log parsing, NA is the balanced Pareto winner across all three hardware platforms. This is consistent with the broader pattern, but notable given that log parsing is also the task where prompt strategy has the largest effect: few-shot prompting raises parsing accuracy from 3.4\% to 54.3\%, making it a more impactful lever than agentic configuration for this workload.
For log analysis, SA is the most energy- and latency-efficient configuration, whereas NA achieves the highest average F1-score.
Given the uniformly low absolute F1-scores for this task (5.5–7.5\%), the practical differences between configurations are small, and the choice between NA and SA is unlikely to affect overall system performance materially.
We synthesize these per-task patterns into concrete design guidelines in Section~\ref{sec:discussion}.

\begin{tcolorbox}[boxsep=2pt,left=2pt,right=2pt,top=2pt,bottom=2pt]
\paragraph{\textbf{RQ4 Conclusion.}} 
Lightweight configurations dominate the efficiency frontier: 55\% of Pareto-optimal configurations use NA and 35\% use SA, while MA accounts for only 2\%. MA is therefore not supported as a default design choice once energy and latency are considered alongside accuracy, even though it improves average vulnerability-detection accuracy. The most effective design lever is not agentic complexity but task-specific model and prompt selection. These findings translate into a clear guideline: start with the simplest configuration (NA), move to heavier workflows only when they provide a measured task-specific benefit, match the model to the task, and use prompt examples where the task demands output structure.
\end{tcolorbox}

\section{Discussion}
\label{sec:discussion}

This section synthesizes the findings of RQ1--RQ4 across tasks, configurations, models, and hardware platforms, and turns them into actionable design guidelines for sustainable LLM-based developer tools. Finally, we discuss the practical implications for tool builders and researchers.

\subsection{Insights across RQs}
\label{subsec:cross_rq}

Four patterns emerge consistently across the four research questions. Together they reframe the role of each design lever in agentic LLM systems for software engineering.

\paragraph{Agentic configuration is a first-order factor, but its effect is asymmetric.}
Agentic configuration is one of the largest single drivers of system behaviour in our study, but its effect on accuracy is not monotonic. On accuracy, NA obtains the highest mean score for four of the five tasks, while MA is the strongest configuration for vulnerability detection. Adding agents otherwise leaves accuracy roughly unchanged (code generation, technical debt detection, log analysis) or degrades it (log parsing, and DA for vulnerability detection). On cost, the same axis grows steeply: averaged across tasks, hardware, models, and prompts, MA consumes 6.36$\times$ as much energy and runs 6.07$\times$ as long as NA, with worst-case slowdowns of up to 160$\times$ for individual task--hardware pairs. The asymmetry is therefore that accuracy gains from additional agents are rare and task-specific, while the resource cost is systematic. The Pareto analysis in RQ4 makes this concrete, showing that out of 66 Pareto-optimal configurations across all task--hardware pairs, 36 are NA, 23 are SA, 6 are DA, and only 1 is MA. This does not mean that one should in general not use agents, but that there should be a clear justification for using them.
Our findings complement the recent work of Tran and Kiela \cite{tran2026single}, who show that under matched reasoning-token budgets, single-agent systems consistently match or outperform multi-agent systems on multi-hop reasoning tasks. 
While their study focuses on reasoning under equal compute budgets and explains this behavior through an information-theoretic perspective, our results reveal a similar pattern across a broader range of software engineering tasks under realistic deployment conditions while explicitly quantifying the associated energy and latency costs. 
Across five software engineering tasks, additional agents provide only limited, task-specific accuracy improvements while imposing systematic increases in energy consumption and inference latency. The two studies therefore converge on the same practical conclusion from different perspectives: agentic complexity should not be treated as a default design choice but adopted only when there is clear evidence of a task-specific benefit.

\begin{figure*}[t]
  \centering
  \includegraphics[width=\textwidth]{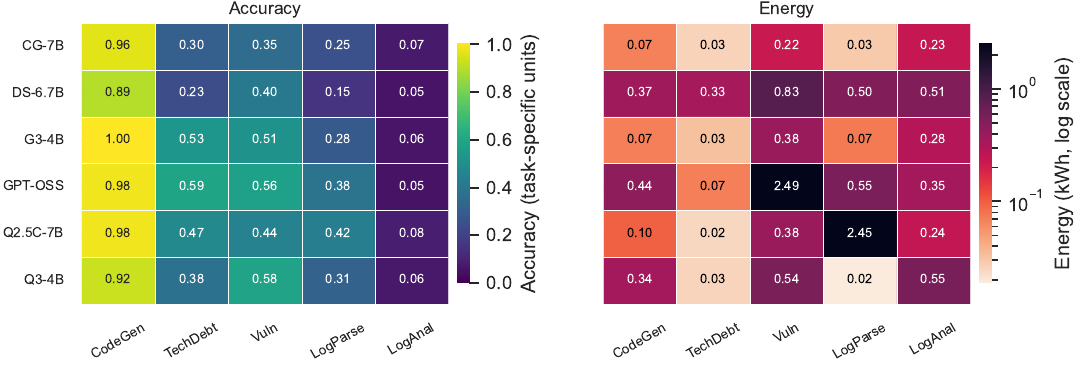}
   \vspace*{-1.20em}
  \caption{Cross-task comparison of LLM effectiveness and energy cost. Rows denote CodeGemma-7B (CG-7B), DeepSeek-Coder-6.7B (DS-6.7B), Gemma-3-4B (G3-4B), GPT-OSS-20B (GPT-OSS), Qwen2.5-Coder-7B (Q2.5C-7B), and Qwen3-4B (Q3-4B); columns denote code generation (CodeGen), technical debt detection (TechDebt), vulnerability detection (Vuln), log parsing (LogParse), and log analysis (LogAnal). The left heatmap reports mean task-specific performance as a proportion---Pass@1, Macro F1, F1-score, or parsing accuracy, depending on the task---and the right heatmap reports mean total pipeline energy per run in kWh. Hardware is first averaged within each model--prompt combination, after which values are averaged over agent configurations and prompting strategies. The energy color scale is logarithmic, whereas cell annotations report untransformed mean values. GPT-OSS values are based on Server-SMU because that model was not evaluated on the other platforms. The differing performance and energy leaders across tasks show no model is uniformly optimal.}
  \label{fig:model_impact_heatmap}
\end{figure*}

\paragraph{Model and prompt are task-specific levers, not universal ones.}
While agentic configuration has a mostly monotonic cost effect, model and prompt have heterogeneous effects whose direction depends on the task. RQ1 shows that no single model is best across all tasks (Pass@1 is maximized by G3-4B, Macro F1 for technical debt by GPT-OSS, vulnerability F1 by Q3-4B, and the two log metrics by Q2.5C-7B), and that prompt strategy can swing log-parsing accuracy from 3.4\% to 54.3\%. RQ2 shows that the lowest-energy model also varies by task, and that the energy gap between models is often larger than the gap between hardware platforms for the same configuration. Figure~\ref{fig:model_impact_heatmap} makes the absence of a universal winner visually explicit: the best accuracy model varies by task, with Q2.5C-7B leading both log tasks, while the lowest-energy model also varies, with CG-7B recurring across several tasks. Model and prompt are therefore the levers that actually deliver task-specific performance improvements, but they require per-task tuning rather than a single global default.

\paragraph{Hardware mostly amplifies the cost of heavier configurations.}
Hardware is the only factor that does not have a strong direct effect on accuracy: across the hardware-balanced subset, accuracy varies by less than 3 percentage points between the three platforms. On the cost side, hardware effects are modest for NA and SA but widen substantially for DA and MA: MA on Workstation-SMU is 3.4$\times$ slower than MA on Server-STF, and the energy gap follows the same direction. Hardware selection therefore becomes a first-order concern primarily as a function of the agentic configuration it is asked to run; it amplifies rather than counteracts the agentic cost.

\paragraph{Energy and latency move together; accuracy moves separately.}
RQ3 shows a near-perfect log--log correlation between energy and latency ($r = 0.987$), driven by the fact that local inference on the same hardware draws roughly constant power. As a consequence, energy savings and latency savings are largely co-achieved, where a configuration change that reduces one will almost always reduce the other. Accuracy, in contrast, is decoupled from this axis---it depends on whether the chosen model and prompt are appropriate for the task, not on how much energy the configuration consumes.

\subsection{Design Guidelines for Sustainable Agentic LLM Systems}
\label{subsec:guidelines}

Drawing on the insights above, we offer the following guidelines for practitioners building LLM-based developer tools. Each guideline is based on our RQ findings.

\paragraph{G1: Start non-agentic, escalate only with evidence.}
For all five tasks studied, NA is the best or tied-for-best configuration on at least one of accuracy, energy, or latency, and it dominates the Pareto front (55\% of all Pareto-optimal configurations). Adding agents should be treated as a hypothesis to test, and not a default. SA is a reasonable next step when a measurable task-specific benefit is observed; DA and MA should be reserved for tasks, such as vulnerability detection in our results, where there is concrete evidence that multi-round reasoning helps enough to justify the cost.

\paragraph{G2: Match the model to the task before tuning the architecture.}
Model choice has a larger and more task-specific effect than agentic configuration. Table~\ref{tab:rq1_model} can be used as a starting point: G3-4B for code generation, GPT-OSS or G3-4B for technical debt detection, Q3-4B for vulnerability detection, Q2.5C-7B for log tasks where accuracy is paramount, and CG-7B/Q3-4B when energy efficiency is important. The attractiveness of a model depends on whether it is sufficient for the task, and not on its parameter count alone: smaller is not automatically more efficient (e.g., Q2.5C-7B can consume more than two orders of magnitude more energy than the best model on log parsing).

\paragraph{G3: Use few-shot prompting for output-structure-sensitive tasks.}
Few-shot prompting is the single most impactful lever for log parsing, raising parsing accuracy from 3.4\% to 54.3\% \emph{and} reducing energy by 24--41\%. It also reduces energy for technical debt detection. For code generation, however, few-shot examples can hurt both accuracy and energy. Few-shot prompting is therefore appropriate when the task is sensitive to output format, and should be tested empirically otherwise.

\paragraph{G4: Choose hardware based on workload, not on raw capability.}
For lightweight (NA/SA) workloads, hardware differences are modest, and choices can be driven by availability or cost. For heavier (DA/MA) workloads, hardware becomes a first-order concern: high-end servers can be 3.4$\times$ faster than workstations on MA, and the energy gap widens accordingly. If the workload must be agentic, invest in suitable hardware; otherwise, the savings from a simpler architecture exceed the savings from upgrading hardware.

\paragraph{G5: Report energy alongside accuracy.}
The strong correlation between energy and latency (RQ3) and the dominance of lightweight configurations on the Pareto front (RQ4) only become visible because all four dimensions were measured uniformly. We encourage authors of new agentic systems to publish accuracy together with energy 
on a fixed reference platform, so that the community can compare designs on a more complete basis than accuracy alone.

\subsection{Implications}
\label{subsec:implications}

\paragraph{For tool builders.}
The strongest practical takeaway is that competitive developer tools can be built around lightweight LLM configurations, provided that the model and prompt are matched to the task. Multi-agent orchestration should be added selectively, instrumented with energy or latency reporting, and validated against a non-agentic baseline before being shipped to users.

\paragraph{For researchers.}
Many existing comparisons of agentic systems report only accuracy and only on a single task. Our results show that conclusions drawn from such single-axis evaluations may be misleading: configurations that look attractive on accuracy alone (e.g., MA on vulnerability detection) can be substantially more expensive than lower-cost alternatives, while configurations that look attractive on cost alone may be insufficient for tasks where the accuracy gap matters. We advocate for multi-objective, multi-task evaluation as a standard practice for agentic LLM research.

\paragraph{For sustainability assessment.}
The energy footprint of agentic LLM systems is not determined by any single design choice, but by the multiplicative interaction of model, agent configuration, prompt, and hardware. Across our 1{,}840 experimental runs, the campaign consumed approximately 651.7~kWh of electricity, which CodeCarbon translates to roughly 110.8~kg~CO$_2$e at the default carbon intensity used in our experiments (mean implied intensity 0.170~kg~CO$_2$e/kWh). The absolute number is modest---comparable to a few hundred kilometres of driving an average passenger car~\cite{usepa}---but it is dominated by a small subset of configurations: the most expensive single configuration alone accounts for about 5.5~kg~CO$_2$e, while the median configuration is roughly two orders of magnitude lower. This means that emissions reduction in agentic LLM workflows is not a question of small efficiency gains across the board, but more about avoiding the worst configurations. We therefore caution against treating CO$_2$ as a stand-alone indicator: under a fixed carbon-intensity assumption, it is a linear transformation of energy and adds no analytical insight beyond the energy results, but its translation into emissions depends strongly on \emph{when} and \emph{where} a model is run. Sustainability assessments of agentic systems should report energy at a fixed reference platform (so that designs can be compared) together with a transparent CO$_2$ estimate that makes the carbon-intensity assumption explicit, and they should highlight the long tail of high-cost configurations rather than the mean.

\section{Threats to Validity}
\label{sec:validity} 

\subsection{Internal Validity} 

Threats to internal validity concern whether the observed effects can be attributed to the experimental factors we studied. To counter such threats, we used fixed prompts and datasets, set the temperature to $0$, disabled agent-level caching, restarted the Ollama server between runs, and executed repeated trials under the same protocol. Still, some sources of stochasticity may remain, such as low-level nondeterminism due to GPU kernels, model-serving behavior, and hardware state. Our three repeated runs per configuration (except for GPT-OSS) capture some variability, but do not fully represent long-term variation due to system temperature, background services, 
or hardware aging. Results involving GPT-OSS should therefore be interpreted with additional caution, since this model was only evaluated on one hardware platform and without repeated runs. To reduce the risk that hardware constraints affected the accuracy analysis, we have done balanced hardware comparisons for RQ1 by averaging the numbers across hardware platforms before averaging across experimentation configurations, preventing that the models run on multiple hardware platforms are weighted more than the GPT-OSS model in the reported numbers.

\subsection{External Validity} 

External validity concerns the generalizability of our findings. We evaluate five different software engineering tasks, which gives a broader coverage than single-task benchmarks, but it is still only a subset of developer workflows. The results may not generalize to tasks such as code review, test generation, repository-level maintenance, or autonomous development. The generalizability is also restricted by the datasets, since real-life use cases generally involve more complex, noisy data and larger inputs compared to the curated benchmark datasets used to evaluate each configuration in this study. Additionally, the datasets used in this study cover a limited set of programming languages: HumanEval targets Python, and the MLCQ and PrimeVul datasets consist of Java and C/C++ code, respectively. The findings on code generation, technical debt detection, and vulnerability detection may therefore not generalize to other languages, particularly those with different syntactic structures, ecosystem conventions, or levels of LLM training coverage.

The model and hardware choices also present a threat to external validity, as we have evaluated only small, local LLMs with a maximum of 20 billion parameters. Results may differ for larger models, and particularly frontier models. Publicly hosted frontier models, however, do not allow direct local measurement of energy consumption, which was one of the main aims of this study. Similarly, cloud deployments may behave differently because of batching, virtualization, datacenter power usage effectiveness, cooling infrastructure, and location-specific carbon intensity.


Our results are affected by our specific implementation of the agentic configurations and the prompts used for the various tasks. Agentic systems can be designed in a multitude of ways, and small details can create significant changes. For simpler, more structured tasks such as log parsing and technical debt detection, we adopted generic orchestration patterns (Actor–Critic and User Proxy–Actor–Critic–Refiner) that are broadly applicable across tasks. For tasks requiring deeper domain reasoning (vulnerability detection, log analysis, and code generation), we defined task-specific agent roles informed by prior literature, allowing us to evaluate representative and validated multi-agent setups rather than purely generic ones. More specialized or extensively tuned multi-agent designs could yield different accuracy results for individual tasks, particularly for vulnerability detection where our MA configuration already shows a meaningful accuracy advantage. Our claim is therefore not that the specific configurations studied are optimal, but that they represent realistic design choices a practitioner would make. The finding that lightweight configurations dominate the Pareto front should be read as evidence about the cost–benefit ratio of typical agentic orchestration, not as a proof that no multi-agent design can ever achieve a better accuracy–energy trade-off. Future work could 
explore the multi-agent design space for individual tasks to identify whether more task-specialized orchestration can shift the Pareto frontier.

\subsection{Construct Validity} 
Construct validity concerns whether the measurements used in the study capture the concepts we intend to analyze. For task performance, we use the term accuracy as an umbrella term, while the actual metric is task-specific: Pass@1 for code generation, Macro F1 and per-class F1 for technical debt detection, F1-score for vulnerability detection and log analysis, and parsing accuracy for log parsing. This improves the fit between metric and task, but it also means that performance values are not directly interchangeable across tasks. Cross-task averages should therefore be interpreted as summaries of relative behavior across the benchmark suite, not as a single universal measure of software engineering capability.

The technical debt task is a particular construct-validity risk because plain accuracy can hide poor performance on minority smell classes. We mitigate this by using Macro F1 as the primary task metric and by reporting per-class F1 scores. However, these metrics still reduce a complex code-quality judgment to a fixed-label classification problem and do not capture severity, maintainability impact, or disagreement among human reviewers.

Energy consumption and latency are also affected by specific measurement choices. CodeCarbon estimates process-level energy consumption from CPU, GPU, and memory signals rather than measuring wall-socket energy with external instrumentation. This provides a consistent basis for comparing configurations, but it may result in greater uncertainty, depending on the idle power draw on various hardware platforms.

A further construct-validity consideration concerns the post-processing used to extract predictions from LLM outputs. Since LLMs generate free-form natural language responses, each task required a task-specific parser or extraction procedure to map the generated text to the expected evaluation format (e.g., labels, code, or structured log templates). To mitigate this issue, our prompts explicitly instructed the models to produce responses in predefined structured formats. Nevertheless, adherence to these formats varied across models, reflecting differences in response style and instruction-following behavior. Deviations from the expected format occurred in some cases, particularly in multi-agent configurations where the final output resulted from multiple intermediate reasoning steps.
Although the post-processing rules were applied consistently across all configurations for a given task, alternative extraction strategies or more robust output normalization could yield different measured accuracy, particularly when a response deviated from the expected format while still conveying the correct semantic answer. Consequently, the reported accuracy reflects both the underlying model performance and the effectiveness of the response extraction procedure adopted in this study.

\section{Conclusion}
\label{sec:conclusion}

This paper evaluated the effectiveness, energy consumption, and latency of agentic LLM configurations across five software engineering tasks, six LLMs, three hardware platforms, and two prompt strategies. The results show that increasing agentic complexity does not provide a general accuracy benefit. The multi-agent configuration improves average accuracy for vulnerability detection, but NA or SA is sufficient, and often strongest, for the other tasks.

The sustainability cost of additional agents is much more systematic. Averaged across tasks, models, prompts, and hardware platforms, MA consumes 6.36$\times$ as much energy and runs 6.07$\times$ as long as NA, while appearing in only one of the 66 Pareto-optimal configurations. Energy and latency are also tightly coupled ($r=0.987$ in log--log space), which means that inefficient configurations are usually both more energy-intensive and slower. Across the full experimental campaign, the runs consumed 651.7~kWh and corresponded to 110.8~kg~CO$_2$e under the fixed carbon-intensity assumption used in the study.

The main practical implication is that sustainable agentic LLM system design should start from a lightweight baseline and escalate only when a task-specific benefit is measured. Model and prompt selection often matter more than adding agents: model choice varies sharply by task, and few-shot prompting is especially important for structured-output tasks such as log parsing. Future evaluations of agentic software engineering systems should therefore report accuracy together with energy and latency, so that claimed improvements can be assessed as full cost--benefit trade-offs rather than accuracy gains in isolation. Beyond evaluation methodology, future work should investigate alternative role assignments and coordination strategies within multi-agent configurations to determine whether more task-specialized designs can shift the Pareto frontier without proportionally increasing energy and latency costs. Future work will also extend the evaluation to repository-scale and tool-augmented software engineering tasks, additional model families, and production deployment environments to assess the generalizability of the observed trade-offs. We will further investigate adaptive orchestration policies that activate additional agents only when task uncertainty or expected accuracy gains justify the energy and latency overhead.




\begin{acks}
The work has been conducted as part of the ENFIELD project (101120657) and the INTEND project (101135576) funded by the European Commission within the HEU Programme.
\end{acks}

\bibliography{main}
\bibliographystyle{ACM-Reference-Format}

\newpage
\appendix

\section{Technical Debt Per-Class Breakdowns}
\label{app:td_per_class}

This appendix reports the per-smell-class F1 breakdowns for the Technical Debt Detection task. The aggregate per-class results are summarized in Table~\ref{tab:rq1_td_per_class_overall} in Section~\ref{subsec:RQ1}; here we provide the same decomposition by agentic configuration (Table~\ref{tab:rq1_td_per_class_agent}), LLM model (Table~\ref{tab:rq1_td_per_class_model}), and prompt strategy (Table~\ref{tab:rq1_td_per_class_prompt}).

\begin{table*}[ht]
\centering
\caption{RQ1 TD-specific per-class results: F1 by smell class and agent configuration (mean $\pm$ std, averaged over models, hardware, and prompts).}
\label{tab:rq1_td_per_class_agent}
\scriptsize
\begin{tabular}{lr@{ $\pm$ }lr@{ $\pm$ }lr@{ $\pm$ }lr@{ $\pm$ }l}
\toprule
Smell Class & \multicolumn{2}{c}{NA} & \multicolumn{2}{c}{SA} & \multicolumn{2}{c}{DA} & \multicolumn{2}{c}{MA} \\
\cmidrule(lr){2-3} \cmidrule(lr){4-5} \cmidrule(lr){6-7} \cmidrule(lr){8-9}
 & mean & std & mean & std & mean & std & mean & std \\
\midrule
No Smell & \textbf{57.4} & \textbf{10.6} & 57.3 & 13.2 & 57.1 & 15.0 & 51.3 & 18.7 \\
Blob & \textbf{32.8} & \textbf{24.6} & 31.3 & 21.3 & 28.6 & 18.7 & 27.4 & 18.7 \\
Data Class & 41.1 & 26.2 & 39.2 & 25.9 & 40.2 & 28.0 & \textbf{41.7} & \textbf{27.1} \\
Feature Envy & 4.9 & 5.8 & 7.5 & 6.1 & \textbf{10.1} & \textbf{7.1} & 9.8 & 8.7 \\
Long Method & 65.0 & 22.6 & \textbf{66.1} & \textbf{19.5} & 61.0 & 23.3 & 58.9 & 23.1 \\
\bottomrule
\end{tabular}
\end{table*}

\begin{table*}[ht]
\centering
\caption{RQ1 TD-specific per-class results: F1 by smell class and LLM model (mean $\pm$ std, averaged over agent configurations, hardware, and prompts).}
\label{tab:rq1_td_per_class_model}
\scriptsize
\begin{tabular}{lr@{ $\pm$ }lr@{ $\pm$ }lr@{ $\pm$ }lr@{ $\pm$ }lr@{ $\pm$ }lr@{ $\pm$ }l}
\toprule
Smell Class & \multicolumn{2}{c}{CG-7B} & \multicolumn{2}{c}{DS-6.7B} & \multicolumn{2}{c}{G3-4B} & \multicolumn{2}{c}{GPT-OSS} & \multicolumn{2}{c}{Q2.5C-7B} & \multicolumn{2}{c}{Q3-4B} \\
\cmidrule(lr){2-3} \cmidrule(lr){4-5} \cmidrule(lr){6-7} \cmidrule(lr){8-9} \cmidrule(lr){10-11} \cmidrule(lr){12-13}
 & mean & std & mean & std & mean & std & mean & std & mean & std & mean & std \\
\midrule
No Smell & 63.7 & 11.7 & 55.8 & 2.2 & \textbf{69.6} & \textbf{6.6} & 67.7 & 2.1 & 47.5 & 9.7 & 38.3 & 14.6 \\
Blob & 2.7 & 3.9 & 17.3 & 6.8 & 53.7 & 10.3 & \textbf{61.2} & \textbf{12.1} & 38.5 & 7.0 & 27.7 & 13.2 \\
Data Class & 2.9 & 7.9 & 14.3 & 6.5 & 57.0 & 3.8 & \textbf{75.3} & \textbf{3.9} & 64.1 & 9.4 & 53.0 & 5.3 \\
Feature Envy & \textbf{13.1} & \textbf{9.2} & 8.0 & 4.8 & 2.5 & 1.7 & 0.0 & 0.0 & 10.7 & 6.0 & 8.8 & 7.4 \\
Long Method & 66.3 & 5.1 & 21.3 & 7.4 & 80.4 & 2.4 & \textbf{88.7} & \textbf{2.6} & 75.5 & 5.0 & 61.6 & 6.8 \\
\bottomrule
\end{tabular}
\end{table*}

\begin{table*}[ht]
\centering
\caption{RQ1 TD-specific per-class results: F1 by smell class and prompt strategy (mean $\pm$ std, averaged over agent configurations, models, and hardware).}
\label{tab:rq1_td_per_class_prompt}
\scriptsize
\begin{tabular}{lr@{ $\pm$ }lr@{ $\pm$ }l}
\toprule
Smell Class & \multicolumn{2}{c}{Zero-shot} & \multicolumn{2}{c}{Few-shot} \\
\cmidrule(lr){2-3} \cmidrule(lr){4-5}
 & mean & std & mean & std \\
\midrule
No Smell & \textbf{56.9} & \textbf{13.8} & 54.6 & 15.6 \\
Blob & \textbf{32.2} & \textbf{21.7} & 27.9 & 19.8 \\
Data Class & 39.9 & 27.8 & \textbf{41.3} & \textbf{25.3} \\
Feature Envy & \textbf{10.1} & \textbf{8.3} & 6.0 & 5.3 \\
Long Method & 61.7 & 22.7 & \textbf{63.8} & \textbf{21.6} \\
\bottomrule
\end{tabular}
\end{table*}

\end{document}